\documentclass[aps,prl,reprint,preprintnumbers,amsmath,amssymb,longbibliography]{revtex4-2}

\pdfoutput=1

\usepackage[dvipsnames]{xcolor}
\usepackage[
    breaklinks=true,
    colorlinks=true,
    urlcolor=blue,
    citecolor=blue,
    linkcolor=blue,
    linktocpage=true
]{hyperref}

\usepackage{bm}
\usepackage{mathtools}
\usepackage{physics}
\usepackage{tikz}
\usepackage{enumitem}

\graphicspath{{fig/}}

\newcommand{\lettersection}[1]{\par\textit{#1}---\ignorespaces}

\makeatletter
\newcommand{\sectionstarNoContents}[1]{%
    \let\sm@savedaddcontentsline\addcontentsline
    \let\sm@savedaddtocontents\addtocontents
    \let\addcontentsline\@gobblethree
    \let\addtocontents\@gobbletwo
    \section*{#1}%
    \let\addcontentsline\sm@savedaddcontentsline
    \let\addtocontents\sm@savedaddtocontents
}
\newcommand{\subsectionstarNoContents}[1]{%
    \let\sm@savedaddcontentsline\addcontentsline
    \let\sm@savedaddtocontents\addtocontents
    \let\addcontentsline\@gobblethree
    \let\addtocontents\@gobbletwo
    \subsection*{#1}%
    \let\addcontentsline\sm@savedaddcontentsline
    \let\addtocontents\sm@savedaddtocontents
}
\newcommand{\bibliographyNoContents}[1]{%
    \let\sm@savedaddcontentsline\addcontentsline
    \let\sm@savedaddtocontents\addtocontents
    \let\addcontentsline\@gobblethree
    \let\addtocontents\@gobbletwo
    \bibliography{#1}%
    \let\addcontentsline\sm@savedaddcontentsline
    \let\addtocontents\sm@savedaddtocontents
}
\makeatother

\usepackage{pgffor}
\foreach \x in {A,B,...,Z,a,b,...,z}{
    \expandafter\xdef\csname cal\x\endcsname{\noexpand\mathcal{\x}}
    \expandafter\xdef\csname scr\x\endcsname{\noexpand\mathscr{\x}}
    \expandafter\xdef\csname bb\x\endcsname{\noexpand\mathbb{\x}}
    \expandafter\xdef\csname rm\x\endcsname{\noexpand\mathrm{\x}}
    \expandafter\xdef\csname it\x\endcsname{\noexpand\mathit{\x}}
    \expandafter\xdef\csname bf\x\endcsname{\noexpand\mathbf{\x}}
    \expandafter\xdef\csname sf\x\endcsname{\noexpand\mathsf{\x}}
    \expandafter\xdef\csname tt\x\endcsname{\noexpand\mathtt{\x}}
    \expandafter\xdef\csname fr\x\endcsname{\noexpand\mathfrak{\x}}
    \expandafter\xdef\csname bm\x\endcsname{\noexpand\bm{\x}}
}

\newcommand{\ii}{\rmi}
\newcommand{\te}{\tau_\text{E}}
\newcommand{\bmell}{\bm{\ell}}
\newcommand{\gm}{\mathrm{GM}}
\newcommand{\gn}{G_{\text{N}}}

\newcommand{\cd}{\mathrm{CD}}
\newcommand{\btz}{\mathrm{BTZ}}
\newcommand{\Tadm}{\calT_{\text{adm}}}
\DeclareMathOperator*{\argmin}{arg\,min}
\DeclareMathOperator{\atanarg}{atan2}

\DeclareMathOperator{\sgn}{sgn}
\DeclareMathOperator*{\Ext}{Min\,Ext}
\DeclareMathOperator{\Arg}{Arg}

\begin{document}

\preprint{YITP-26-65}

\title{Saddle Incompatibility Generates Genuine Multientropy in Heavy Local Quenches}

\author{Kosei Fujiki}
\email{kosei.fujiki@yukawa.kyoto-u.ac.jp}
\affiliation{
    Center for Gravitational Physics and Quantum Information,
    Yukawa Institute for Theoretical Physics, Kyoto University,
	Kitashirakawa Oiwakecho, Sakyo-ku, Kyoto 606-8502, Japan
}

\author{Kenya Tasuki}
\email{kenya.tasuki@yukawa.kyoto-u.ac.jp}
\affiliation{
    Center for Gravitational Physics and Quantum Information,
    Yukawa Institute for Theoretical Physics, Kyoto University,
	Kitashirakawa Oiwakecho, Sakyo-ku, Kyoto 606-8502, Japan
}

\date{\today}

\begin{abstract}
We study the time evolution of tripartite genuine multientropy in heavy local quenches via AdS${}_3$/CFT${}_2$.
Its linear response around the vacuum cancels pointwise for arbitrary adjacent intervals.
At finite backreaction, the vacuum-subtracted genuine multientropy arises only from an incompatibility gap between the triplet of windings selected by the trivalent geodesic network and the windings selected independently by the bipartite geodesics.
The heavy-light vacuum-block calculation reproduces the same formula from the boundary perspective.
The time dependence is kinematically fixed and is independent of the heavy operator dimension.
These results show that the genuine multipartite entanglement is controlled by global saddle transition rather than by local energy response or quasiparticle propagation.
\end{abstract}

\maketitle

\lettersection{Introduction}
Entanglement entropy is the canonical measure of bipartite entanglement for pure states in quantum many-body systems, quantum field theories, and quantum gravity~\cite{Bombelli:1986rw,Srednicki:1993im}.
In two-dimensional conformal field theories (CFTs), the replica construction gives a universal single-interval vacuum entanglement entropy~\cite{Holzhey:1994we,Calabrese:2004eu,Calabrese:2009qy,Casini:2009sr}.
In holographic CFTs, the AdS/CFT correspondence~\cite{Maldacena:1997re,Gubser:1998bc,Witten:1998qj} maps it to extremal bulk areas through the Ryu--Takayanagi (RT) formula and its generalizations~\cite{Ryu:2006bv,Ryu:2006ef,Lewkowycz:2013nqa,Hubeny:2007xt}.
This geometric description establishes entanglement as a probe of the bulk and underlies the broader perspective that spacetime is encoded in entanglement structure~\cite{Headrick:2007km,Swingle:2009bg,VanRaamsdonk:2010pw,Freedman:2016zud}.

Quantum quenches turn this static dictionary into a dynamical probe by driving the system out of equilibrium~\cite{Calabrese:2005in,Calabrese:2007rg,Calabrese:2007mtj,Balasubramanian:2010ce,Hartman:2013qma,Liu:2013iza,Liu:2013qca,Ugajin:2013xxa,Nozaki:2013wia,Nozaki:2014hna,Asplund:2014coa,Caputa:2014eta,Rangamani:2015agy,Calabrese:2016xau,Numasawa:2016emc,Rottoli:2024ylt}.
The resulting growth of bipartite entanglement often admits a quasiparticle description in which entangled excitations propagate ballistically~\cite{Calabrese:2005in,Calabrese:2007rg,Rottoli:2024ylt}.
For infinitesimal perturbations, the first law of entanglement equates the entropy response with the variation of modular energy and leads holographically to the linearized Einstein equations~\cite{Blanco:2013joa,Lashkari:2013koa,Bhattacharya:2013bna,Faulkner:2013ica}.

Nevertheless, entanglement entropy alone does not determine how quantum correlations are shared among three or more subsystems~\cite{Greenberger:1989tfe,Vidal:1998re,Coffman:1999jd,Linden:1999qve,Dur:2000zz,Plenio:2007zz,Horodecki:2009zz}.
Holographic entropy vectors and mixed-state entanglement measures obey strong constraints, revealing a highly constrained multipartite entanglement structure in states dual to classical gravity~\cite{Hayden:2011ag,Bao:2015bfa,Bao:2018gck,Takayanagi:2017knl,Nguyen:2017yqw,Umemoto:2019jlz,Akers:2019gcv,Hayden:2021gno,Mori:2025gqe,Balasubramanian:2025hxg,Balasubramanian:2026chr}.
Since a quench redistributes correlations across multiple regions, multipartite entanglement naturally probes the resulting dynamics.
Indeed, heavy local quenches exhibit rich mixed-state entanglement dynamics~\cite{Kusuki:2019rbk,Kusuki:2019evw}, while multipartite entanglement signals show qualitatively distinct behavior in coarse-grained late-time regimes of global quenches~\cite{Balasubramanian:2025jhq}.

Multientropy (ME) extends the replica construction symmetrically to multipartite systems and is dual to an extremal ramified soap film~\cite{Gadde:2022cqi,Penington:2022dhr}.
Genuine multientropy (GM) subtracts lower-partite contributions to isolate the genuinely multipartite component~\cite{Harper:2024ker,Liu:2024ulq,Iizuka:2025ioc,Iizuka:2025caq}.
For a local excitation in a free scalar CFT${}_2$, the raw ME admits a quasiparticle-pair interpretation that leads to vanishing GM~\cite{Harper:2024ker}.
This contrast raises the question of whether multipartite entanglement dynamics in strongly coupled holographic CFTs is likewise governed by the preceding descriptions or instead reveals a different mechanism.

In this Letter, we determine the tripartite GM dynamics after a heavy local operator quench in AdS${}_3$/CFT${}_2$ and identify global saddle incompatibility as its governing mechanism.
In the bulk calculation, the linear response of GM cancels pointwise for arbitrary adjacent intervals at all times, although every constituent entropy responds.
After minimizing the trivalent network over its absolute windings and the three RT geodesics independently over their relative windings, the vacuum-subtracted GM~\eqref{eq:bulk-excess-gm} at finite backreaction is nonnegative.
It becomes strictly positive precisely when the minimized trivalent network cannot realize the three independently minimized RT geodesics.
In the heavy-light CFT calculation, the same compatibility gap~\eqref{eq:cft-excess-gm} is reproduced through the mismatch between one common tripartite monodromy branch and the three independently minimizing bipartite branches.
Bipartite saddle transitions determine the support in the time direction, whereas tripartite network transitions create cusps.
In the sharp-quench limit, the profile~\eqref{eq:gm-excess-sharp-quench} is fixed kinematically independent of the heavy operator dimension.
These results establish global saddle incompatibility and transitions as a distinct organizing principle for multipartite entanglement dynamics in holographic CFTs.

\lettersection{Holographic genuine multientropy}
In defining the R\'{e}nyi-$n$ $q$-partite ME $S_n^{(q)}$, one introduces $n^{q-1}$ replicas of the state $\ket{\Phi}$, labelled by $\bmr = (r_1, r_2, \dots, r_{q-1}) \in (\bbZ_n)^{q-1}$.
For $i=1,\dots,q-1$, the indices of subsystem $A_i$ are contracted cyclically along the $i$-th replica direction $r_i$, while those of the last subsystem $A_q$ are contracted within each replica.
Let $P_n^{(q)}$ denote the permutation operator that implements such contractions.
The R\'{e}nyi-$n$ $q$-partite ME is defined as
\begin{align}
    S_n^{(q)}(\{A_i\}) &\coloneqq \frac{1}{1-n} \frac{1}{n^{q-2}} \log{Z[\Sigma_n^{(q)}(\{A_i\})]},\\
    Z[\Sigma_n^{(q)}(\{A_i\})] &\coloneqq \Tr[P_n^{(q)}(\{A_i\}) (\ketbra{\Phi})^{\otimes n^{q-1}}].
    \label{eq:RenyiMEdef}
\end{align}
We write $S^{(q)} \coloneqq \lim_{n\to1} S^{(q)}_n$ for the von Neumann limit $n\to1$ studied below.

\begin{figure}[t]
    \centering
    \includegraphics[width=0.49\textwidth]{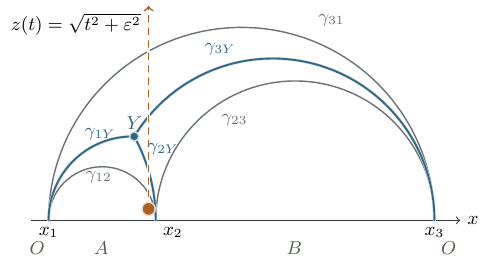}
    \caption{Network geometry in Poincar\'{e} coordinates.
    Gray RT geodesics bound the entanglement wedge polygon.
    Blue trivalent-leg geodesics meet at $Y$ in its interior.
    The dashed red arrow is the projection of the falling-particle trajectory.}
    \label{fig:network-geometry}
\end{figure}

We use the holographic soap-film prescription in which ME is dual to the extremal area of a ramified $q$-valent surface $\Gamma^{(q)}$ homologous to the partition~\cite{Gadde:2022cqi,Penington:2022dhr}.
For each admissible topology, the surface is extremized, and the least-area admissible extremal surface is selected as
\begin{equation}
    S^{(q)}(A_1:\cdots:A_q) = \Ext_{\Gamma^{(q)}} \frac{\text{Area}(\Gamma^{(q)})}{4\gn},
\end{equation}
where we set the AdS radius to unity.
The case $q=2$ reduces to the RT formula, while in AdS${}_3$/CFT${}_2$ the case $q=3$ is represented by a trivalent network, as in Fig.~\ref{fig:network-geometry}.

The lower-partite contributions are subtracted from ME to define the genuine multientropy (GM)~\cite{Iizuka:2025ioc,Iizuka:2025caq}.
For $q=3$~\cite{Harper:2024ker,Liu:2024ulq,Iizuka:2025ioc,Iizuka:2025caq},
\begin{equation}
    \gm^{(3)} \coloneq S^{(3)}(A:B:O) - \frac{1}{2} \sum_{X = A,B,O} S^{(2)}(X:X^c).
\end{equation}
Tripartite GM is a multipartite entanglement signal, namely a local-unitary invariant that vanishes for layerwise-separable states~\cite{Ju:2026xfh,Gadde:2026msg}.
A nonzero value detects genuine tripartite quantum correlations beyond the layerwise bipartite structure.
For adjacent intervals $A=[x_1,x_2]$, $B=[x_2,x_3]$, and $O=(A\cup B)^c$ with $x_1 < x_2 < x_3$ in CFT${}_2$, the vacuum GM is UV-finite and independent of the interval lengths~\cite{Harper:2024ker,Harper:2025uui}.
In AdS${}_3$/CFT${}_2$, it is computed as the difference between the trivalent-network length and half the sum of the three RT-geodesic lengths,
\begin{equation}
    \left. \gm^{(3)} \right|_{\text{vacuum}} = \frac{3}{4\gn} \log{\frac{2}{\sqrt{3}}}.
\end{equation}

\lettersection{Holographic heavy local operator quench}
The quench is generated at the origin $x=0$ by a heavy operator $\psi$ of chiral weights $h_\psi = \bar{h}_\psi$ with $h_\psi/c$ fixed in the large-central-charge limit $c \to \infty$.
The Euclidean quench width $\varepsilon > 0$ places $\psi$ at $w_0 = -i\varepsilon$ and its conjugate $\psi^\dagger$ at $w_\infty = +i\varepsilon$ in the complex coordinate $w = x + i\te$.
The setup and conventions are summarized in \hyperref[app:setup]{Appendix~A}.
The unnormalized state is $\ket{\psi(t)} = e^{-iHt} e^{-\varepsilon H} \psi(x=0) \ket{0}$, where $H$ is the CFT Hamiltonian and $\ket{0}$ is the vacuum.
We define $\alpha \coloneqq \sqrt{1-24h_\psi/c}$ and $M \coloneqq 1-\alpha^2 = 24h_\psi/c$.
We present the Letter in the conical-defect regime $0 < \alpha < 1$, while the continuation to the BTZ black hole regime follows by analytic continuation $\alpha \mapsto i\nu$ with $\nu \coloneqq \sqrt{24h_\psi/c - 1} > 0$~\cite{SM}.
The sharp-quench limit takes $\varepsilon \to 0$ away from light-cone and saddle-transition singularities.

The bulk dual is a massive particle falling in AdS${}_3$ with trajectory $z^2 - t^2 = \varepsilon^2$, $x = 0$ in Poincar\'{e} coordinates $(t,x,z)$~\cite{Nozaki:2013wia,Asplund:2014coa}.
Mapping the trajectory to the origin, the backreacted spacetime is AdS${}_3$ with conical identification $\theta \sim \theta + 2\pi \alpha$,
\begin{equation}
    \dd{s}^2 = \dd{\rho}^2 - \cosh^2{\rho} \dd{\tau}^2 + \sinh^2{\rho} \dd{\theta}^2.
    \label{eq:conical-defect-metric}
\end{equation}
The global coordinates $(\tau, \theta, \rho)$ are related to the Poincar\'{e} coordinates $(t,x,z)$ by a $t$-dependent Poincar\'{e}-to-global map collected in \hyperref[app:setup]{Appendix~A}.

For any observable $Q$ computed in the quench, we define its vacuum-subtracted value as $\varDelta{Q} \coloneq Q - \left. Q \right|_{\text{vacuum}}$.
In particular, $\varDelta \gm^{(3)}$ is the dynamical excess of tripartite GM studied below.

\lettersection{Pointwise cancellation of the linear response}
Let $\delta_M Q$ denote the term linear in $M$ around the vacuum, induced by the metric perturbation $\delta_M g$.
Let $\gamma_{[a,b]}^{[0]}$ denote the vacuum geodesic anchored at the boundary points $a<b$, and let $\gamma_{P_- P_+}^{[0]} \subset \gamma_{[a,b]}^{[0]}$ be an arc with actual endpoints $P_\pm = (x_\pm, y_\pm)$ with $x_- < x_+$.
The metric contribution to its first-order length variation is
\begin{equation}
    \delta_M \calL_{\gamma_{P_- P_+}^{[0]}} = \int_{x_-}^{x_+} \dd{x} \, K(x,t) W[a, b](x),
\end{equation}
where the kernel and the weighting function are given by
\begin{align}
    K(x,t) &\coloneqq \frac{M\varepsilon^2}{[(t-x)^2 + \varepsilon^2]^2} + \frac{M\varepsilon^2}{[(t+x)^2 + \varepsilon^2]^2}\\
    W[a,b](x) &\coloneqq \frac{(b-x)(x-a)}{b-a}.
\end{align}
Here, $K(x,t)$ encodes the contribution of the holographic stress tensor, whose profile is localized around the ballistic rays $x=\pm t$ with width $\varepsilon$, consistent with the quasiparticle picture.

Write $x_1 = x_2 - L$ and $x_3 = x_2 + R$ with $L,R>0$.
The vacuum trivalent vertex has coordinates
\begin{align}
    x_Y &= x_2 + \frac{LR(L-R)}{2(L^2 + LR + R^2)},\\
    z_Y &= \frac{\sqrt{3}}{2} \frac{LR(L+R)}{L^2+LR+R^2}.
\end{align}
Each leg $\gamma_{iY}^{[0]}$ from $x_i$ to $Y$ is an arc of a full semicircle whose other boundary endpoint is
\begin{equation}
    \bar{x}_i \coloneqq x_Y + \frac{z_Y^2}{x_Y - x_i}.
\end{equation}
For $L>R$, the actual projected leg ranges are $[x_1,x_Y]$, $[x_2,x_Y]$, and $[x_Y,x_3]$.
Writing $W_{ij} \coloneqq W[x_i,x_j]$, $W_{i\bar{i}} \coloneqq W[x_i,\bar{x}_i]$, and $W_{\bar{i}i} \coloneqq W[\bar{x}_i,x_i]$, direct polynomial expansion gives the pointwise identities
\begin{align}
    W_{1\bar{1}} &= \frac{1}{2} (W_{12} + W_{13}), \quad x_1 \leq x \leq x_2,\\
    W_{1\bar{1}} + W_{2\bar{2}} &= \frac{1}{2} (W_{13} + W_{23}), \quad x_2 \leq x \leq x_Y,\\
    W_{\bar{3}3} &= \frac{1}{2} (W_{13} + W_{23}), \quad x_Y \leq x \leq x_3.
\end{align}
Since the unperturbed network is extremal, displacement of $Y$ does not contribute at first order.
Thus the linear response of the trivalent geodesic network is canceled by the linear response of the RT geodesics,
\begin{equation}
    \delta_M \gm^{(3)} = \frac{1}{4\gn} \sum_{i=1}^3 \delta_M \calL_{\gamma_{iY}^{[0]}} - \frac{1}{8\gn} \sum_{(i,j)} \delta_M \calL_{\gamma_{ij}^{[0]}} = 0.
\end{equation}
The case $L<R$ follows by reflection and the case $L=R$ by continuity.
This cancellation is pointwise in $(x,t)$, and holds for arbitrary adjacent tripartitions.
Thus GM removes the local linear response, though each constituent entropy responds at this order.
Any nonzero value must arise from a mechanism beyond local energy response.

\lettersection{Global saddle incompatibility at finite backreaction}
For two boundary points $i$ and $j$ in the conical-defect geometry~\eqref{eq:conical-defect-metric}, each RT geodesic connecting them is labeled by a relative winding $m_{ij} \in \bbZ$ and has length
\begin{align}
    \calL_{ij}(m_{ij}) &= \log{\calI_{ij}(m_{ij})} + \rho_i + \rho_j - \log{2},\\
    \calI_{ij}(m) &\coloneqq \cos{(\tau_i - \tau_j)} - \cos{(\theta_i - \theta_j + 2\pi \alpha m)}.
\end{align}
Products and sums run over the cyclically oriented pairs $(ij|X)=(12|A),(23|B),(31|O)$.
Each subsystem $(ij|X)$ admits two choices of relative winding $m_{ij}$ for its RT geodesic, corresponding to whether or not the associated entanglement wedge contains the conical defect.
For the ordering $0 < x_2 < -x_1 < x_3$ used below, the sets $\calD_{ij|X}$ of allowed relative windings are
\begin{equation}
    \calD_{12|A}=\{0,1\},\quad
    \calD_{23|B}=\{0,-1\},\quad
    \calD_{31|O}=\{0,-1\}.
    \label{eq:pair-domains}
\end{equation}

A trivalent network must instead use a triplet $T = [(\ell_1, \ell_2, \ell_3)] \in \bbZ^3/\bbZ(1,1,1)$ of absolute windings defined up to a common shift.
It induces a tuple $\bmm(T)=(m_{12}(T), m_{23}(T), m_{31}(T))$ of relative windings $m_{ij}(T) \coloneqq \ell_i - \ell_j$.
After extremizing with respect to the trivalent vertex as discussed in \hyperref[app:extremization]{Appendix~C}~\cite{Anegawa:2025prn}, the network length for a fixed $T$ is
\begin{equation}
    \calL_{\Gamma_T} = \frac{1}{2} \sum_{(i,j)} \log{\calI_{ij}(m_{ij}(T))} + \sum_i \rho_i + \frac{3}{2} \log{\frac{2}{3}}.
\end{equation}
The admissible domain $\Tadm$ consists of those absolute winding classes for which all three geodesics are spacelike and homologous to their respective subsystems~\cite{SM}.
There are three possible absolute winding classes $T$, distinguished by which of the three faces of the bulk partition defined by the trivalent network contains the conical defect.
For the ordering above, the three candidates are
\begin{align}
    T_A &= [(0,0,0)], & \bmm(T_A) &= (0,0,0),\\
    T_B &= [(0,-1,0)], & \bmm(T_B) &= (1,-1,0),\\
    T_O &= [(1,0,0)], & \bmm(T_O) &= (1,0,-1),
\end{align}
where $T_X$ corresponds to the case in which the conical defect lies within the face containing subsystem $X$.

\begin{figure}[t]
    \centering
    \includegraphics[width=0.49\textwidth]{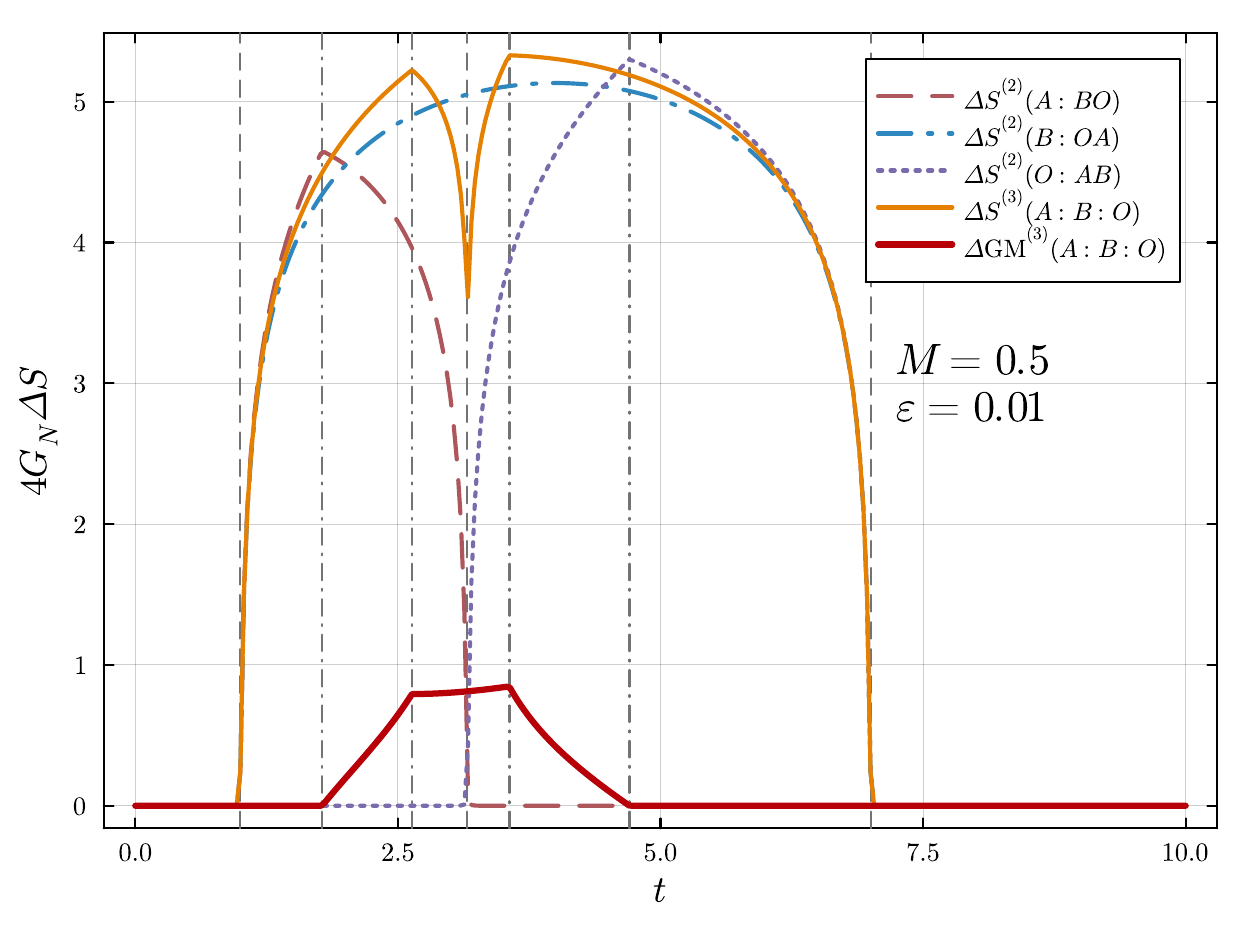}
    \caption{Time evolution of entropies  at $\varepsilon=0.01$ and $M=0.5$.
    Endpoints are $(x_1,x_2,x_3)=(-\sqrt{10}, 1, 7)$.
    The vacuum-subtracted GM is supported between RT-saddle transitions and develops cusps at trivalent-network saddle transitions.}
    \label{fig:plt_con_btz}
\end{figure}

\begin{figure*}[t]
    \centering
    \includegraphics[width=0.99\textwidth]{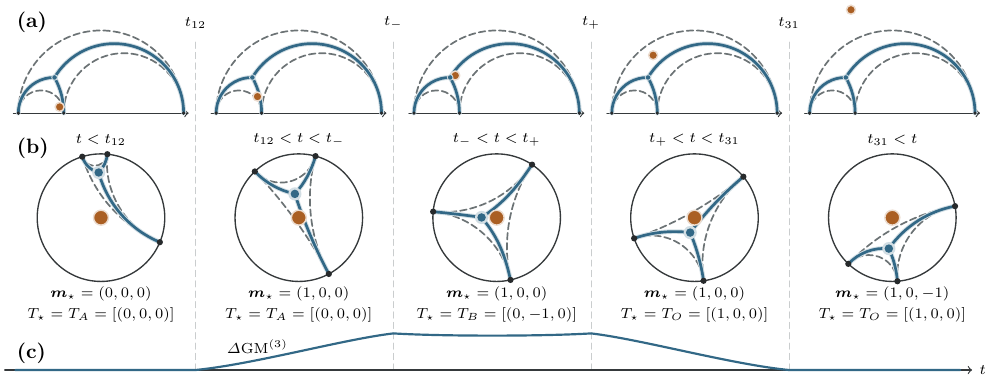}
    \caption{Sharp-quench branch sequence.
    (a) Poincar\'{e} falling-particle geometry for $0<x_2<-x_1<x_3$.
    (b) Corresponding conical-defect geometry centered on the particle.
    Dashed RT geodesics bound the entanglement wedge polygon, and blue trivalent-leg geodesics meet in its interior.
    (c) Schematic profile of vacuum-subtracted GM.}
    \label{fig:transition-sequence}
\end{figure*}

After vacuum subtraction, the radial and constant terms cancel.
Independently minimizing the three bipartite entropies over their respective admissible domains, while minimizing the trivalent network over $\Tadm$, yields
\begin{equation}
    \varDelta{\gm^{(3)}} = \frac{1}{8\gn} \log{\frac{\displaystyle\min_{T\in\Tadm} \prod_{(i, j)} \calI_{ij}(m_{ij}(T))}{\displaystyle\prod_{(ij|X)} \min_{m \in \calD_{ij|X}} \calI_{ij}(m)}} \geq 0.
    \label{eq:bulk-excess-gm}
\end{equation}
Every $T\in\Tadm$ in the numerator supplies an admissible candidate to the three minimizations in the denominator, proving the nonnegativity of the vacuum-subtracted GM.
The inequality is strict if and only if no admissible network induces a winding tuple that independently minimizes all three bipartite entropies.
The RT saddle transitions determine the temporal support of the excess, whereas trivalent-network transitions produce its cusps, as shown in Fig.~\ref{fig:plt_con_btz}.
Fig.~\ref{fig:transition-sequence} summarizes the sequence of bipartite and trivalent branches analyzed below.

The minimization can be performed analytically in the sharp-quench limit, with all endpoint orderings classified in \hyperref[app:classification]{Appendix~D}~\cite{SM}.
For ordering $0 < \varepsilon \ll x_2 < -x_1 < x_3$, the RT saddle transition times are
\begin{equation}
    t_{12} \coloneqq \sqrt{|x_1|x_2}, \qquad t_{31} \coloneqq \sqrt{|x_1|x_3}.
\end{equation}
The trivalent-network transition times are
\begin{align}
    t_- &\coloneqq \sqrt{\frac{x_2[2|x_1| x_3 - x_2(x_3 - |x_1|)]}{2x_2 + x_3 - |x_1|}},\\
    t_+ &\coloneqq \sqrt{\frac{x_3[2|x_1| x_2 + x_3(|x_1| - x_2)]}{x_2 + 2x_3 - |x_1|}}.
\end{align}
In this limit, we obtain
\begin{equation}
    \lim_{\varepsilon \to 0} \varDelta{\gm^{(3)}} = \frac{1}{8\gn} \log{\calF(t)},
    \label{eq:gm-excess-sharp-quench}
\end{equation}
where
\begin{equation}
    \calF(t) \coloneqq
    \begin{cases}
        1, & 0 < t < t_{12}, \\
        \displaystyle \frac{(t-x_2)(t+|x_1|)}{(|x_1|-t)(t+x_2)}, & t_{12} < t < t_-, \\
        \displaystyle \frac{(t+x_2)(t+x_3)}{(t-x_2)(x_3-t)}, & t_- < t < t_+, \\
        \displaystyle \frac{(t+|x_1|)(x_3-t)}{(t-|x_1|)(t+x_3)}, & t_+ < t < t_{31}, \\
        1, & t_{31} < t.
    \end{cases}
\end{equation}
The same expression holds in the BTZ black hole regime above the threshold~\cite{SM}.
All factors depending on $h_\psi$ cancel in this sharp-quench limit expression, leaving a function determined solely by the kinematics.
This universality contrasts with the static heavy-state result, where tripartite GM depends on the horizon radius or the conformal dimension and develops volume-law scaling~\cite{Anegawa:2025prn}.

\lettersection{Compatibility gap from CFT}
The same compatibility gap follows in a large-central-charge CFT${}_2$ with a sparse light spectrum and vacuum Virasoro block dominance.
Define the orbifold theory operator $\Psi_n^{(q)} \coloneqq \psi^{\otimes n^{q-1}}$.
Let $g_A$ and $g_B$ be the cyclic permutations of the replicas along the first and second label directions, respectively.
The endpoint permutations are $(g_1, g_2, g_3) = (g_A, g_A^{-1}g_B, g_B^{-1})$.
The normalized five-point replica correlator for the tripartite R\'{e}nyi ME is
\begin{equation}
    Z_n^{(3)} = \frac{\expval{(\Psi_n^{(3)})^\dagger(w_\infty) \prod_{i=1}^3 \calT_{g_i}^{(3)}(w_i) \Psi_n^{(3)}(w_0)}}{\expval{(\Psi_n^{(3)})^\dagger(w_\infty) \Psi_n^{(3)}(w_0)}},
\end{equation}
where the twist operators $\calT_g^{(q)}$ have chiral weights $h_n^{(q)} = \bar{h}_n^{(q)} = (cn^{q-2}/24) (n-1/n)$ and are treated as light near the von Neumann limit $n \to 1$.

The heavy-light correlator is obtained by the uniformization of the heavy background~\cite{Asplund:2014coa,Fitzpatrick:2014vua,Fitzpatrick:2015zha,Perlmutter:2015iya,Alkalaev:2015fbw} as
\begin{align}
    Z_n^{(3)} &\simeq C_n^{(3)} \abs{\frac{\zeta_{12} \zeta_{23} \zeta_{31}}{w_{12} w_{23} w_{31}}}^{2h_n^{(3)}} \abs{\frac{u_1' u_2' u_3'}{u_{12} u_{23} u_{31}}}^{2h_n^{(3)}},\\
    w &\mapsto \zeta(w) = \frac{w - w_0}{w - w_\infty} \mapsto u(\zeta) = \zeta^\alpha.
    \label{eq:uniformization-map}
\end{align}
Here $w_{ij} \coloneqq w_i - w_j$, $\zeta_{ij} \coloneqq \zeta_i - \zeta_j$, $u_{ij} \coloneqq u_i - u_j$, and $u_i' \coloneqq \eval{(\dd{u}/\dd{\zeta})}_{\zeta=\zeta_i}$ and $C_n^{(3)}$ is the twist OPE coefficient.
Applying the uniformization to the bipartite correlators gives the vacuum-subtracted bipartite entropies~\cite{Asplund:2014coa}.
For $(ij|X) = (12|A), (23|B), (31|O)$, we obtain
\begin{align}
    \varDelta{S}^{(3)} &= \frac{c}{6} \log{\abs{\frac{u_{12} u_{23} u_{31}}{\zeta_{12} \zeta_{23} \zeta_{31} u_1' u_2' u_3'}}},\\
    \varDelta{S}^{(2)}(X:X^c) &= \frac{c}{6} \log{\abs{\frac{u_{ij}^2}{\zeta_{ij}^2 u_i' u_j'}}}.
\end{align}

Since the uniformization map~\eqref{eq:uniformization-map} is multivalued, each entropy must be minimized independently over the branches of the map.
After Lorentzian continuation, the M\"obius map sends the endpoints to phases as
\begin{equation}
    \zeta_i = e^{\ii \phi_i} = \frac{w_i - w_0}{w_i - w_\infty}, \quad \bar{\zeta}_i = e^{-\ii \bar{\phi}_i} = \frac{\bar{w}_i - \bar{w}_0}{\bar{w}_i - \bar{w}_\infty}.
\end{equation}
Its multi-valued lifts are described by
\begin{align}
    \varphi_i &= \phi_i + 2\pi n_i, \quad \bar{\varphi}_i = \bar{\phi}_i + 2\pi\bar{n}_i, \quad n_i,\bar{n}_i \in \bbZ,\\
    \alpha \varphi_i &= \theta_i + \tau_i + 2\pi\alpha n_i, \quad \alpha \bar{\varphi}_i = \theta_i - \tau_i + 2\pi\alpha \bar{n}_i.
\end{align}
so that $u_i = e^{\ii \alpha \varphi_i}$ and $\bar{u}_i = e^{-\ii \alpha \bar{\varphi}_i}$.
We choose diagonal branches $n_i=\bar{n}_i$ such that the corresponding CFT saddle is real.
For each bipartite correlator, the branch is specified by the relative winding
$m_{ij}=n_i-n_j$, whereas the tripartite correlator requires one common
absolute winding class $T=[(n_1,n_2,n_3)]$.
The corresponding branch-dependent factors therefore satisfy
\begin{align}
    \eval{ \abs{\frac{u_{ij}^2}{u_i' u_j'}}^2 }_{m_{ij}} &= \left( \frac{2}{\alpha^2} \right)^2 [\calI_{ij}(m_{ij})]^2,\\
    \eval{ \abs{\frac{u_{12} u_{23} u_{31}}{u_1' u_2' u_3'}}^2 }_T &= \left( \frac{2}{\alpha^2} \right)^3 \prod_{(i,j)}\calI_{ij}(m_{ij}(T)).
\end{align}
Minimizing each entropy with respect to the branches in its own admissible domain gives
\begin{equation}
    \varDelta{\gm^{(3)}} = \frac{c}{12} \log{\frac{\displaystyle\min_{T\in\Tadm} \prod_{(i, j)} \calI_{ij}(m_{ij}(T))}{\displaystyle\prod_{(ij|X)} \min_{m \in \calD_{ij|X}} \calI_{ij}(m)}}.
    \label{eq:cft-excess-gm}
\end{equation}
The Brown--Henneaux relation $c = 3/(2\gn)$~\cite{Brown:1986nw} makes Eq.~\eqref{eq:cft-excess-gm} agree with Eq.~\eqref{eq:bulk-excess-gm}.
The boundary calculation matches not only the final profile but also the optimization architecture.
Three independently minimized bipartite branches compete with one common tripartite branch, so bulk winding and CFT branch encode the same compatibility obstruction.

\lettersection{Discussion and outlook}
The bulk-boundary agreement identifies the global saddle incompatibility as the source of nonzero vacuum-subtracted GM.
It becomes positive precisely when no triplet of independently minimizing RT geodesics admits a common realization as one admissible trivalent network.

In the sharp quench limit, the temporal support coincides with the interval during which the falling particle lies inside the associated entanglement wedge polygon (EWP)~\cite{Fujiki:2026ucr}, bounded by the three RT geodesics.
The particle enters and exits the EWP at the RT saddle transitions and crosses the interior trivalent-leg geodesics at the trivalent-network saddle transitions that generate the cusps.
Tripartite GM therefore provides a dynamical probe of EWP interior geometry, which is relevant for multipartite entanglement information.

The compatibility gap mechanism for genuine multipartite entanglement in chaotic, strongly coupled holographic CFT is distinct from quasiparticle descriptions of bipartite entanglement dynamics or their generalization to multipartite entanglement~\cite{Berthiere:2024sio}.
It also complements membrane descriptions of bipartite entanglement and multipartite entanglement signals, which are valid in coarse-grained late-time and large-region dynamics in chaotic systems~\cite{Mezei:2018jco,Balasubramanian:2025jhq}.
Our results show that the compatibility gap provides a sharp, all-time, fine-grained description of multipartite entanglement dynamics that these descriptions do not capture.

Analytic control relies on solvable trivalent network extremization and conformally fixed kinematics.
Disjoint intervals and higher-partite systems add cross ratios, homology sectors, competing block channels, and multiple junctions.
Higher-dimensional setups, other quench protocols~\cite{Hartman:2013qma,Calabrese:2007mtj,Ugajin:2013xxa}, and other related quantities~\cite{Yuan:2024yfg,Iizuka:2025elr,Basak:2024uwc,Basak:2026seh} provide additional tests of the mechanism.
At quantum order, quantum extremal surfaces can determine whether corrections shift or smooth the classical transitions~\cite{Faulkner:2013ana,Engelhardt:2014gca}.
These extensions will test how broadly the compatibility gap organizes multipartite entanglement dynamics.

\lettersection{Acknowledgments}
The authors thank Tadashi Takayanagi and Kotaro Tamaoka for discussions.
This work is supported by MEXT KAKENHI Grant-in-Aid for Transformative Research Areas (A) through the ``Extreme Universe'' collaboration: Grant Number~21H05182.
KF is supported by Grant-in-Aid for JSPS Fellows No.~26KJ1554.
KT is supported by Grant-in-Aid for JSPS Fellows No.~25KJ1455.

\bibliographyNoContents{ref}

\begin{center}
    {\large\bfseries END MATTER}
\end{center}

\begin{figure}[b]
    \centering
    \includegraphics[width=0.49\textwidth]{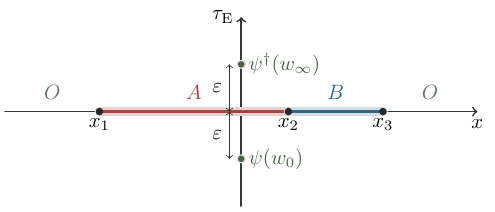}
    \caption{Subsystem and Euclidean regulator setup for the heavy local quench.
    The adjacent intervals are $A=[x_1,x_2]$, $B=[x_2,x_3]$, and $O=(A\cup B)^c$.
    Twist operators $\calT_{g_i}$ are inserted at the endpoints $x_i$, while heavy operators $\psi$ and $\psi^\dagger$ are inserted at $w_0 = -i\varepsilon$ and $w_\infty = +i\varepsilon$, respectively.}
    \label{fig:setup}
\end{figure}

\sectionstarNoContents{Appendix~A: Quench and Coordinate Conventions}
\label{app:setup}
The adjacent intervals are $A=[x_1,x_2]$, $B=[x_2,x_3]$ and $O=(A\cup B)^c$.
The Euclidean insertions are $\psi(w_0)$ at $w_0 = -i\varepsilon$ and $\psi^\dagger(w_\infty)$ at $w_\infty = +i\varepsilon$ in the complex coordinate $w = x + i\te$.
See Fig.~\ref{fig:setup}.

The dual particle follows $x=0$ and $z^2 - t^2 = \varepsilon^2$ in Poincar\'{e} coordinates $(t,x,z)$.
The map from Poincar\'{e} coordinates to the static frame is
\begin{align}
    \tilde{\tau} &= \atanarg(z^2 + x^2 - t^2 + \varepsilon^2, 2\varepsilon t),\\
    \tilde{\theta} &= \atanarg(z^2 + x^2 - t^2 - \varepsilon^2, 2\varepsilon x),\\
    r &= \frac{1}{z} \sqrt{x^2 + \frac{(z^2 + x^2 - t^2 - \varepsilon^2)^2}{4\varepsilon^2}}.
\end{align}
Here $\atanarg(a,b) \coloneqq \Arg(a + \ii b) \in (-\pi,\pi]$.
The static metric is
\begin{equation}
    \dd{s}^2 = - (r^2 + 1-M) \dd{\tilde{\tau}}^2 + \frac{\dd{r}^2}{r^2 + 1-M} + r^2 \dd{\tilde{\theta}}^2.
\end{equation}
With $M=1-\alpha^2$, $\tau = \alpha \tilde{\tau}$, $\theta = \alpha \tilde{\theta}$, and $r = \alpha \sinh{\rho}$, the static metric becomes Eq.~\eqref{eq:conical-defect-metric} with $\theta \sim \theta + 2\pi\alpha$.
The time-dependent endpoint data $(\tau_i, \theta_i, \rho_i)$ are obtained by evaluating the map at the regulated boundary endpoints.

\sectionstarNoContents{Appendix~B: Admissible Windings}
A candidate RT geodesic is a spacelike geodesic homologous to the corresponding subsystem.
For an oriented subsystem $(ij|X)$, the domain $\calD_{ij|X}$ consists of the candidates $m_{ij}$ of relative windings satisfying
\begin{equation}
    \abs{\theta_i - \theta_j + 2\pi \alpha m_{ij}} < 2\pi \alpha, \quad \calI_{ij}(m_{ij}) > 0.
\end{equation}
The three RT geodesics are minimized independently over
\begin{equation}
    \bmm = (m_{12}, m_{23}, m_{31}) \in \calD_{12|A} \times \calD_{23|B} \times \calD_{31|O}.
\end{equation}

A trivalent network instead uses a triplet of absolute windings subject to the homology condition and to the requirement that the three geodesics meet at a common bulk vertex.
A common shift leaves all relative windings unchanged, so a physical network branch is an equivalence class
\begin{equation}
    T = [\bmell] \in \bbZ^3/\bbZ(1,1,1), \quad \bmell = (\ell_1, \ell_2, \ell_3) \in \bbZ^3.
\end{equation}
Its cyclic projection is
\begin{equation}
    \pi(T) \coloneqq \bmm(T) = (m_{12}(T), m_{23}(T), m_{31}(T)),
\end{equation}
with components $m_{ij}(T) \coloneqq \ell_i - \ell_j$ satisfying
\begin{equation}
    m_{12}(T) + m_{23}(T) + m_{31}(T) = 0.
\end{equation}
The admissible domain $\Tadm$ consists of those equivalence classes for which the associated geodesics satisfy the spacelike, homology, and noncrossing conditions~\cite{SM}.
Consequently, $\pi(\Tadm) \subset \calD_{12|A} \times \calD_{23|B} \times \calD_{31|O}$ is a subset of the product of the three independent bipartite domains.

\sectionstarNoContents{Appendix~C: Trivalent Network Extremization}
\label{app:extremization}
We use the embedding space formalism in which AdS${}_3$ is represented by the hyperboloid $X\cdot X=-1$ in $\bbR^{2,2}$~\cite{Anegawa:2025prn}.
For a fixed winding class $T=[\bmell]$, define $\vartheta_i(t) \coloneqq \theta_i(t) + 2 \pi \alpha \ell_i$ and the boundary null vectors
\begin{equation}
    P_i(T; t) = (\cos{\tau_i}, \sin{\tau_i}, \cos{\vartheta_i}, \sin{\vartheta_i}).
\end{equation}
The boundary vectors are null and satisfy
\begin{equation}
    P_i \cdot P_j = - \cos{(\tau_i - \tau_j)} + \cos{(\vartheta_i - \vartheta_j)} = - \calI_{ij}(m_{ij}(T)).
\end{equation}

For a regulated endpoint at radial coordinate $\rho_i$, the length of the geodesic leg ending at the trivalent vertex $Y$ is
\begin{equation}
    \calL_{iY} = \log{[\frac{e^{\rho_i}}{2} (-2P_i \cdot Y)]}.
\end{equation}
The constrained network length is therefore
\begin{equation}
    \calL(T; Y) = \sum_{i=1}^3 \log{[\frac{e^{\rho_i}}{2} (-2 P_i \cdot Y)]} + \lambda (Y \cdot Y + 1),
\end{equation}
where $\lambda$ is the Lagrange multiplier.
Variation with respect to $Y$ gives
\begin{equation}
    \sum_{i=1}^3 \frac{P_i}{(-2P_i \cdot Y)} = \lambda Y.
\end{equation}
Contracting the stationarity equation with $Y$ and using $P_i\cdot Y=-A_i/2$ and $Y\cdot Y=-1$ yields $\lambda=3/2$.

For cyclic $(i,j,k)$, contraction with $P_i$ gives
\begin{equation}
    \frac{\calI_{ij}(m_{ij}(T))}{(-2P_i \cdot Y)(-2P_j \cdot Y)} + \frac{\calI_{ki}(m_{ki}(T))}{(-2P_k \cdot Y)(-2P_i \cdot Y)} = \frac{3}{4}.
\end{equation}
The three cyclic equations imply
\begin{equation}
    (-2 P_i \cdot Y)^2 = \frac{8}{3} \frac{\calI_{ij}(m_{ij}(T)) \calI_{ki}(m_{ki}(T))}{\calI_{jk}(m_{jk}(T))}.
\end{equation}
Substitution into the physical network length gives the extremal network length
\begin{equation}
    \calL(T) = \frac{1}{2} \sum_{(i,j)} \log{\calI_{ij}(m_{ij}(T))} + \sum_i \rho_i + \frac{3}{2} \log{\frac{2}{3}}.
\end{equation}

\sectionstarNoContents{Appendix~D: Transition Classification in Sharp-Quench Limit}
\label{app:classification}
In the conical-defect regime, take the sharp-quench limit $\varepsilon \to 0$.
For $x_1 < 0 < x_2 < x_3$, the independently minimizing pair-winding tuple follows
\begin{equation}
    \bmm_\star(t) = \begin{cases}
        (0,0,0), & 0 < t < t_{12},\\
        (1,0,0), & t_{12} < t < t_{31},\\
        (1,0,-1), & t_{31} < t.
    \end{cases}
\end{equation}
The middle tuple violates closure and therefore cannot be the projection of a tuple of absolute windings.
The closure-compatible candidate tuples are
\begin{align}
    T_A &=[(0,0,0)], & \pi(T_A) &= (0,0,0),\\
    T_B &=[(0,-1,0)], & \pi(T_B) &= (1,-1,0),\\
    T_O &=[(0,-1,-1)], & \pi(T_O) &= (1,0,-1).
\end{align}
Reflection about the insertion point reduces the ordering to four cases as shown in Table~\ref{tab:transition}.

\begin{table}[b]
    \centering
    \begin{ruledtabular}
        \begin{tabular}{c c c}
        Case & Endpoint ordering & Trivalent-network sequence\\
        \hline
        O & $0<x_1<x_2<x_3$ & no transition\\
        A-I & $0<-x_1<x_2<x_3$ & $T_A \to T_O$\\
        A-II & $0<x_2<-x_1<x_3$ & $T_A \to T_B \to T_O$\\
        A-III & $0<x_2<x_3<-x_1$ & $T_A \to T_O$
        \end{tabular}
    \end{ruledtabular}
    \caption{Classification of endpoint orderings and trivalent-network branch sequences in the sharp-quench limit.}
    \label{tab:transition}
\end{table}

Case O has zero vacuum-subtracted GM.
Every nontrivial case has support on the interval $t_{12} < t < t_{31}$.
The trivalent-network saddle transition time in cases A-I and A-III is
\begin{equation}
    t_\triangle \coloneqq \sqrt{ \abs{ \frac{x_1 [2x_2x_3 - \abs{x_1}(x_2 + x_3)]}{x_2 + x_3 - 2 \abs{x_1}} } },
\end{equation}
with $t_{12} < t_\triangle < t_{31}$, leading to a triangular profile.
For case A-II, the two trivalent-network saddle transition times are
\begin{align}
    t_- &\coloneqq \sqrt{ \frac{x_2 [2\abs{x_1} x_3 - x_2 (x_3 - \abs{x_1})]}{2x_2 + x_3 - \abs{x_1}} },\\
    t_+ &\coloneqq \sqrt{ \frac{x_3 [2\abs{x_1} x_2 + x_3 (\abs{x_1} - x_2)]}{x_2 + 2x_3 - \abs{x_1}} },
\end{align}
with $t_{12} < t_- < \abs{x_1} < t_+ < t_{31}$, leading to a trapezoidal profile.

Together with the A-II profile in Eq.~\eqref{eq:gm-excess-sharp-quench}, the remaining nontrivial profiles are
\begin{equation}
    \calF_{\text{A-I}}(t) = \begin{cases}
        1, \qquad & t < t_{12},\\
        \displaystyle \frac{(t-\abs{x_1})(t+x_2)}{(t+\abs{x_1})(x_2-t)}, \qquad & t_{12} < t < t_\triangle,\\
        \displaystyle \frac{(t+\abs{x_1})(x_3-t)}{(t-\abs{x_1})(t+x_3)}, \qquad & t_\triangle < t < t_{31},\\
        1, \qquad & t_{31} < t,
    \end{cases}
\end{equation}
and
\begin{equation}
    \calF_{\text{A-III}}(t) = \begin{cases}
        1, \qquad & t < t_{12},\\
        \displaystyle \frac{(\abs{x_1} + t)(t-x_2)}{(\abs{x_1}-t)(t+x_2)}, \qquad & t_{12} < t < t_\triangle,\\
        \displaystyle \frac{(\abs{x_1}-t)(t+x_3)}{(t+\abs{x_1})(t-x_3)}, \qquad & t_\triangle < t < t_{31},\\
        1, \qquad & t_{31} < t.
    \end{cases}
\end{equation}
Case O has $\calF_{\text{O}}(t) = 1$.
The profiles are continuous and satisfy $\calF(t) \geq 1$ throughout their support.

\clearpage

\onecolumngrid

\setcounter{page}{1}

\renewcommand{\thepage}{S\arabic{page}}
\renewcommand{\theequation}{S\arabic{equation}}
\renewcommand{\thefigure}{S\arabic{figure}}
\renewcommand{\thetable}{S\arabic{table}}

\pdfbookmark[0]{Supplementary Material}{supplementary-material}

\begin{center}
    {\large\bfseries SUPPLEMENTAL MATERIAL}
\end{center}

\setcounter{section}{0}
\setcounter{subsection}{0}
\setcounter{equation}{0}
\setcounter{figure}{0}
\setcounter{table}{0}
\setcounter{secnumdepth}{2}

\tableofcontents

\section{Conventions and Setup}
We consider the unnormalized quenched state
\begin{equation}
    \ket{\psi(t)} = e^{-\ii H t} e^{-\varepsilon H} \psi(x=0) \ket{0},
\end{equation}
where $\psi$ is a heavy operator with chiral weights $h_\psi = \bar{h}_\psi$ and $\varepsilon$ is the quench width.
The heavy operator has $h_\psi = \bar{h}_\psi = (c/24)(1-\alpha^2)$, where $\alpha$ is related to the mass parameter $M$ in the bulk by
\begin{equation}
    M = \frac{24h_\psi}{c} = 1 - \alpha^2.
\end{equation}

In complex coordinates $w = x + \ii \te$, the heavy operator $\psi$ is inserted at $w_0 = -\ii \varepsilon$ while its conjugate $\psi^\dagger$ is inserted at $w_\infty = \ii \varepsilon$.
We take adjacent intervals $A = [x_1, x_2]$, $B = [x_2, x_3]$, and $O = (A \cup B)^c$ with $x_1 < x_2 < x_3$.

The bulk is backreacted by the massive particle dual to the heavy operator, with trajectory $z^2 - t^2 = \varepsilon^2$, $x = 0$ in Poincar\'e coordinates $(t, x, z)$.
The transformation to a static frame in which the particle remains at the origin is
\begin{align}
    \tilde{\tau} &= \atanarg(z^2 + x^2 - t^2 + \varepsilon^2, 2 \varepsilon t),\\
    \tilde{\theta} &= \atanarg(z^2 + x^2 - t^2 - \varepsilon^2, 2 \varepsilon x),\\
    r &= \frac{1}{z} \sqrt{x^2 + \frac{(z^2 + x^2 - t^2 - \varepsilon^2)^2}{4 \varepsilon^2}}.
    \label{eq_sm:poincare-global-map}
\end{align}
We set the AdS radius to unity and the metric in the static frame is
\begin{equation}
    \dd{s}^2 = - (r^2 + 1 - M) \dd{\tilde{\tau}}^2 + \frac{\dd{r}^2}{r^2 + 1 - M} + r^2 \dd{\tilde{\theta}}^2.
\end{equation}
For $0 < \alpha < 1$ below the BTZ threshold, the geometry is a conical defect with deficit angle $2\pi(1-\alpha)$.
Above the threshold, we write $\alpha = \ii \nu$ with $\nu = \sqrt{M-1} > 0$.
The resulting geometry is a BTZ black hole with horizon radius $r_h = \nu$.

\section{Perturbative Holographic Analysis}
Here we perturbatively compute the multientropy after the heavy local quench, following the treatment of entanglement entropy in Ref.~\cite{Nozaki:2013wia}.

\subsection{First-order geodesic perturbation}
In the Fefferman--Graham gauge, the backreacted metric near the asymptotic boundary is
\begin{equation}
    \dd{s}^2 = \frac{1}{z^2} \left( \dd{z}^2 + h_{ab} \dd{x^a} \dd{x^b} \right), \qquad h_{ab} = \eta_{ab} + M \tau_{ab} z^2 + \order{M^2}.
\end{equation}
Here $\tau_{ab}$ denotes the first-order Fefferman--Graham coefficient sourced by the heavy local quench.
The component required for our calculation is~\cite{Nozaki:2013wia}
\begin{equation}
    \delta_M g_{xx} = M \tau_{xx} = K(x, t) \coloneqq \frac{M \varepsilon^2}{[(t-x)^2 + \varepsilon^2]^2} + \frac{M \varepsilon^2}{[(t+x)^2 + \varepsilon^2]^2}.
\end{equation}
Let $\gamma_{[a,b]}^{[0]}$ denote the unperturbed vacuum geodesic anchored at $a$ and $b$ where $a<b$.
It is the semicircle
\begin{equation}
    z = Z(x) \coloneqq \sqrt{(x - a)(b - x)}.
\end{equation}
Consider an arc $\gamma_{P_-P_+}^{[0]} \subset \gamma_{[a,b]}^{[0]}$ with endpoints $P_\pm = (Z_\pm, x_\pm)$, where $a \leq x_- < x_+ \leq b$.
Along the arc, the induced metric is
\begin{equation}
    G_{xx}^{[0]} = \frac{1 + Z'^2}{Z^2}, \qquad \delta_M G_{xx} = \delta_M g_{xx} = K(x, t).
\end{equation}
The first-order response of the geodesic length under the metric perturbation is
\begin{align}
    \delta_M {\calL}_{\gamma_{P_-P_+}^{[0]}} &= \frac{1}{2} \int_{\gamma_{P_-P_+}^{[0]}} \dd{x} \sqrt{G^{[0]}} G^{[0]\alpha\beta} \delta_M G_{\alpha\beta}\\
    &= \int_{x_-}^{x_+} \dd{x} K(x, t) W[a, b](x), \qquad W[a, b](x) \coloneqq \frac{(x-a)(b-x)}{b-a}.
\end{align}
The anchors $a, b$ of the full geodesic determine the weight, while the actual arc endpoints $x_\pm$ determine the integration range.
This fixed-branch calculation is analogous to that used to derive the first-law response of entanglement to a small state perturbation~\cite{Blanco:2013joa,Lashkari:2013koa,Bhattacharya:2013bna,Faulkner:2013ica}.

\subsection{Perturbative cancellation in the genuine combination}
Write
\begin{equation}
    x_1 = x_2 - L, \qquad x_3 = x_2 + R, \qquad L>0, \qquad R>0.
\end{equation}
The vacuum trivalent network for the adjacent tripartition is the Steiner tree with the vertex at
\begin{equation}
    x_Y = x_2 + \frac{LR(L-R)}{2(L^2 + LR + R^2)}, \qquad z_Y = \frac{\sqrt{3}}{2} \frac{LR(L+R)}{L^2 + LR + R^2}.
\end{equation}
Each leg from $x_i$ to the vertex $Y$ is an arc of a full semicircle.
Its other boundary anchor lies at the mirror point $\bar{x}_i = x_Y + z_Y^2/(x_Y-x_i)$ and is given explicitly by
\begin{equation}
    \bar{x}_1 = x_2 + \frac{LR}{2L+R}, \qquad \bar{x}_2 = x_2 + \frac{2LR}{L-R}, \qquad \bar{x}_3 = x_2 - \frac{LR}{L+2R}.
\end{equation}

\begin{figure}[t]
    \centering
        \includegraphics[width=0.7\linewidth]{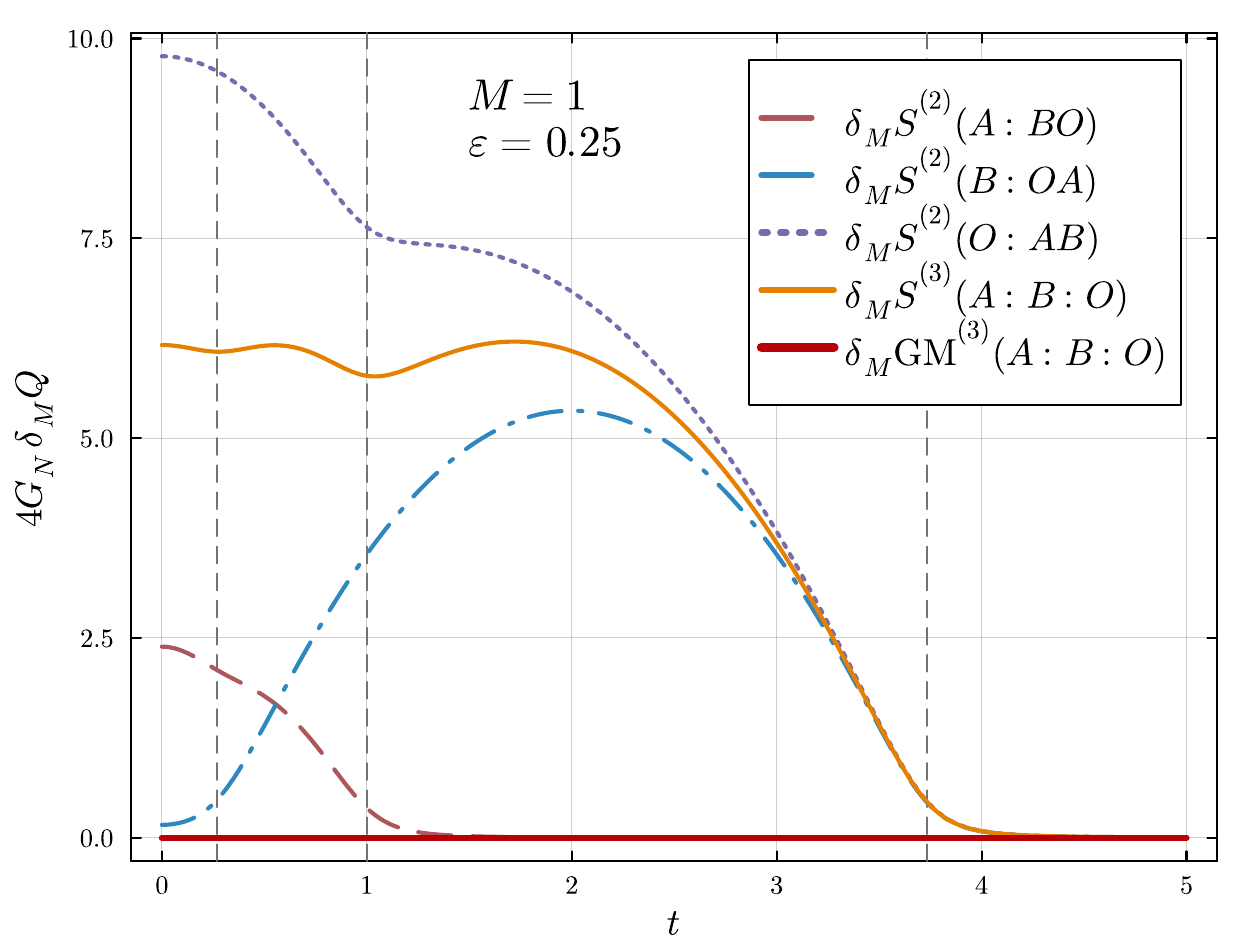}
        \caption{Perturbative holographic entropies for $(x_1,x_2,x_3) = (-1,2-\sqrt{3},2+\sqrt{3})$, $M = 1$, and $\varepsilon = 0.25$.
        The first-order response of the tripartite GM (solid red curve) vanishes at all times.}
    \label{fig_sm:per_plots}
\end{figure}

Set $L > R$ without loss of generality.
The case $L<R$ follows by reflecting the configuration under $L\leftrightarrow R$, and the symmetric case $L=R$ is obtained by continuity, with the geodesic through $x_2$ becoming the vertical geodesic.
The projected ranges of the three legs are $[x_1, x_Y]$, $[x_2, x_Y]$, and $[x_Y, x_3]$.
The first two overlap on $[x_2, x_Y]$, so the combined weight is evaluated piecewise on $[x_1, x_2]$, $[x_2, x_Y]$, and $[x_Y, x_3]$.
Explicit expansion gives the following polynomial identities in the three regions,
\begin{align}
    W[x_1, \bar{x}_1](x) &= \frac{1}{2} \left( W[x_1, x_2](x) + W[x_1, x_3](x) \right), \qquad x_1 \leq x \leq x_2,\\
    W[x_1, \bar{x}_1](x) + W[x_2, \bar{x}_2](x) &= \frac{1}{2} \left( W[x_1, x_3](x) + W[x_2, x_3](x) \right), \qquad x_2 \leq x \leq x_Y,\\
    W[\bar{x}_3, x_3](x) &= \frac{1}{2} \left( W[x_1, x_3](x) + W[x_2, x_3](x) \right), \qquad x_Y \leq x \leq x_3.
\end{align}
The left-hand sides are the local weights of the trivalent leg geodesics, while the right-hand sides are the weights of the RT geodesics for bipartite entropies.
Since the unperturbed network is extremal, the displacement of the vertex does not contribute to the first-order correction.
The pointwise identities show that the first-order contribution to the vacuum-subtracted tripartite GM vanishes for arbitrary adjacent intervals at any time,
\begin{equation}
    \delta_M \gm^{(3)} = \frac{1}{4G_N} \left( \sum_{i=1}^3 \delta_M \calL_{\gamma_{iY}^{[0]}} - \frac{1}{2} \sum_{(i,j)} \delta_M \calL_{\gamma_{ij}^{[0]}} \right) = 0.
\end{equation}
The numerical evaluation in Fig.~\ref{fig_sm:per_plots} confirms this cancellation for a representative configuration.

\section{Fully Back-reacted Holographic Analysis}
In this section, we compute the tripartite GM by geodesic minimization in the fully backreacted geometry.
The saddle-selection problem here differs from the static pure-BTZ setup in Ref.~\cite{Anegawa:2025prn}.
There, the boundary points are compared at fixed global time, while in our local-quench setup, the intervals are fixed at a common Poincar\'e time.
Under the map~\eqref{eq_sm:poincare-global-map} from Poincar\'e to global coordinates, the endpoints generally acquire different global times, so the fully backreacted answer is controlled by both their time-dependent positions and the winding sectors of the geodesics.

\subsection{Conical defect regime below the threshold}
For $0 < \alpha < 1$, set
\begin{equation}
    r = \alpha \sinh{\rho}, \qquad \tau = \alpha \tilde{\tau}, \qquad \theta = \alpha \tilde{\theta}.
\end{equation}
The local metric is global AdS$_3$,
\begin{equation}
    \dd{s}^2 = \dd{\rho}^2 - \cosh^2{\rho} \dd{\tau}^2 + \sinh^2{\rho} \dd{\theta}^2,
\end{equation}
but with the periodicity $\theta \sim \theta + 2\pi\alpha$.

For two boundary points $i$ and $j$, a lifted RT geodesic is labeled by a relative winding $m_{ij}^{(2)} \in \bbZ$ and has length
\begin{align}
    \calL_{ij}^{\cd}(m_{ij}^{(2)}) &= \log{\calI_{ij}^{\cd}(m_{ij}^{(2)})} + \rho_i + \rho_j - \log{2},\\
    \calI_{ij}^{\cd}(m) &\coloneqq \cos(\tau_i - \tau_j) - \cos(\theta_i - \theta_j + 2\pi\alpha m).
\end{align}
A real spacelike RT geodesic requires $\calI_{ij}^{\cd}(m_{ij}^{(2)}) > 0$.
The same embedding invariant controls the bulk-to-boundary segment length used below, after replacing one of the boundary null vectors by the bulk vertex vector.
The bipartite entanglement entropy is obtained by minimizing over admissible $m_{ij}^{(2)}$.
We orient the three boundary arcs cyclically as $(ij|X)=(12|A),(23|B),(31|O)$ and set $m_{ji}^{(2)}=-m_{ij}^{(2)}$.
For an oriented interval $X$, the domain $\calD^{(2)}_{ij|X}$ contains the relative windings whose corresponding RT geodesic is real and homologous to $X$,
\begin{equation}
    0<|\theta_i - \theta_j + 2\pi\alpha m_{ij}^{(2)}| < 2\pi\alpha, \qquad \calI_{ij}^{\cd}(m_{ij}^{(2)}) > 0.
\end{equation}
The tuple of relative windings and its product domain are
\begin{equation}
    \bmm^{(2)} \coloneqq (m_{12}^{(2)}, m_{23}^{(2)}, m_{31}^{(2)}), \qquad \calD^{(2)} \coloneqq \calD_{12|A}^{(2)} \times \calD_{23|B}^{(2)} \times \calD_{31|O}^{(2)}.
    \label{eq_sm:pair-domain}
\end{equation}
Products over $(i,j)$ below follow the cyclic order $(12),(23),(31)$.

For the trivalent network, choose an absolute winding
\begin{equation}
    \bmell = (\ell_1, \ell_2, \ell_3) \in \bbZ^3, \qquad \vartheta_i = \theta_i + 2\pi\alpha \ell_i.
\end{equation}
A physical network branch is the equivalence class $T = [\bmell] \in \bbZ^3/\bbZ(1,1,1)$.
Its cyclic projection is a tuple of relative windings
\begin{equation}
    \pi(T) \coloneqq \bmm^{(3)}(T) = (m_{12}^{(3)}(T), m_{23}^{(3)}(T), m_{31}^{(3)}(T)), \qquad m_{ij}^{(3)}(T) = \ell_i - \ell_j,
\end{equation}
yielding
\begin{equation}
    m_{12}^{(3)}(T) + m_{23}^{(3)}(T) + m_{31}^{(3)}(T) = 0.
\end{equation}
The lifted invariant is
\begin{equation}
    \calI_{ij}^{\cd}(T) \coloneqq \calI_{ij}^{\cd}(m_{ij}^{(3)}(T)) = \cos(\tau_i - \tau_j) - \cos(\vartheta_i - \vartheta_j).
\end{equation}
The finite set $\Tadm$ contains the tuple of absolute windings for which all three invariants are positive and the homology condition holds,
\begin{equation}
    \max_i \vartheta_i - \min_i \vartheta_i < 2\pi\alpha, \qquad \calI_{ij}^{\cd}(m_{ij}^{(3)}) > 0.
\end{equation}
For the projection, we have $\pi(\Tadm) \subset \calD^{(2)}$.
For the domains in Eq.~\eqref{eq_sm:pair-domain}, representatives in the gauge $\ell_1 = 0$ are
\begin{equation}
    T_A = [(0,0,0)], \qquad T_B = [(0,-1,0)], \qquad T_O = [(0,-1,-1)].
\end{equation}
They induce $\bmm^{(3)}(T_A) = (0,0,0)$, $\bmm^{(3)}(T_B) = (1,-1,0)$, and $\bmm^{(3)}(T_O) = (1,0,-1)$.

Now fix $T \in \Tadm$ and extremize the trivalent length over the vertex position $Y$.
We use embedding coordinates in $\mathbb{R}^{2,2}$ with inner product
\begin{equation}
    X \cdot X' \coloneqq -X_{-1}X'_{-1} - X_0X'_0 + X_1X'_1 + X_2X'_2.
\end{equation}
The asymptotic boundary null vector of the lifted endpoint $i$ is
\begin{equation}
    P_i=(\cos\tau_i,\sin\tau_i,\cos\vartheta_i,\sin\vartheta_i),
\end{equation}
and $Y$ is the unit timelike embedding-space vector of the trivalent vertex, $Y^2 = -1$.
We write the trivalent network length functional with the Lagrange multiplier $\lambda$ for the constraint $Y^2=-1$ as
\begin{equation}
    \calL(T; Y) = \sum_{i=1}^3 \log{\left[ \frac{e^{\rho_i}}{2} (-2 P_i \cdot Y)\right]} + \lambda (Y^2 + 1).
\end{equation}
The stationarity condition with respect to the vertex position $Y$ gives
\begin{equation}
    \sum_{i=1}^3 \frac{P_i}{(-2 P_i \cdot Y)} = \lambda Y.
\end{equation}
Contracting this equation with $Y$ fixes $\lambda = 3/2$.
Since
\begin{equation}
    - 2 P_i \cdot P_j = 2\calI_{ij}^{\cd}(T),
\end{equation}
contracting the stationarity condition with $P_j$ gives
\begin{equation}
    (-2 P_j \cdot Y)^2 = \frac{8}{3} \frac{\calI_{ij}^{\cd}(T) \calI_{jk}^{\cd}(T)}{\calI_{ki}^{\cd}(T)}, \qquad (i,j,k) \text{ cyclic}.
\end{equation}
These equations are solved by
\begin{equation}
    Y = \frac{2}{3} \sum_{i=1}^3 \left( \frac{8}{3} \frac{\calI_{ij}^{\cd}(T) \calI_{ki}^{\cd}(T)}{\calI_{jk}^{\cd}(T)} \right)^{-1/2} P_i, \qquad (i,j,k) \text{ cyclic}.
\end{equation}
The extremized trivalent length in a fixed tuple of absolute windings is
\begin{equation}
    \calL_{\Gamma_T} = \frac{1}{2} \log{[\calI_{12}^{\cd}(T) \calI_{23}^{\cd}(T) \calI_{31}^{\cd}(T)]} + \sum_{i=1}^3 \rho_i + \frac{3}{2}\log{\frac{2}{3}}.
\end{equation}
The radial terms cancel between the tripartite ME and the bipartite entanglement entropies in the genuine combination.
At any fixed absolute windings, the remaining constant equals the vacuum GM, so the vacuum-subtracted fixed branch value vanishes exactly.
Minimizing each entropy independently in its own domain gives
\begin{equation}
    \varDelta{\gm}^{(3)} = \frac{1}{8G_N} \log{\frac{\displaystyle\min_{T \in \Tadm} \prod_{(i,j)} \calI_{ij}^{\cd}(T)}{\displaystyle\prod_{(ij|X)} \min_{m^{(2)}_{ij} \in \calD^{(2)}_{ij|X}} \calI_{ij}^{\cd}(m_{ij}^{(2)})}}.
    \label{eq_sm:bulk-excess-gme}
\end{equation}
For later use, define the independent RT and trivalent minimizer sets by
\begin{align}
    \calM^{(2)}(t) &\coloneqq \argmin_{\bmm^{(2)} \in \calD^{(2)}} \prod_{(i,j)} \calI_{ij}^{\cd}(m_{ij}^{(2)}),\\
    \calM^{(3)}(t) &\coloneqq \argmin_{T \in \Tadm} \prod_{(i,j)} \calI_{ij}^{\cd}(T).
\end{align}
The inclusion makes the ratio at least one and proves nonnegativity of the vacuum-subtracted tripartite GM, $\varDelta{\gm}^{(3)} \geq 0$.
The exact equality condition is
\begin{equation}
    \varDelta{\gm}^{(3)} = 0 \iff \pi(\calM^{(3)}(t)) \cap \calM^{(2)}(t) \neq \varnothing.
\end{equation}
The vacuum-subtracted GM is strictly positive if and only if this intersection is empty.

\begin{figure}[t]
    \centering
    \begin{minipage}[t]{0.49\textwidth}
        \centering
        \includegraphics[width=\linewidth]{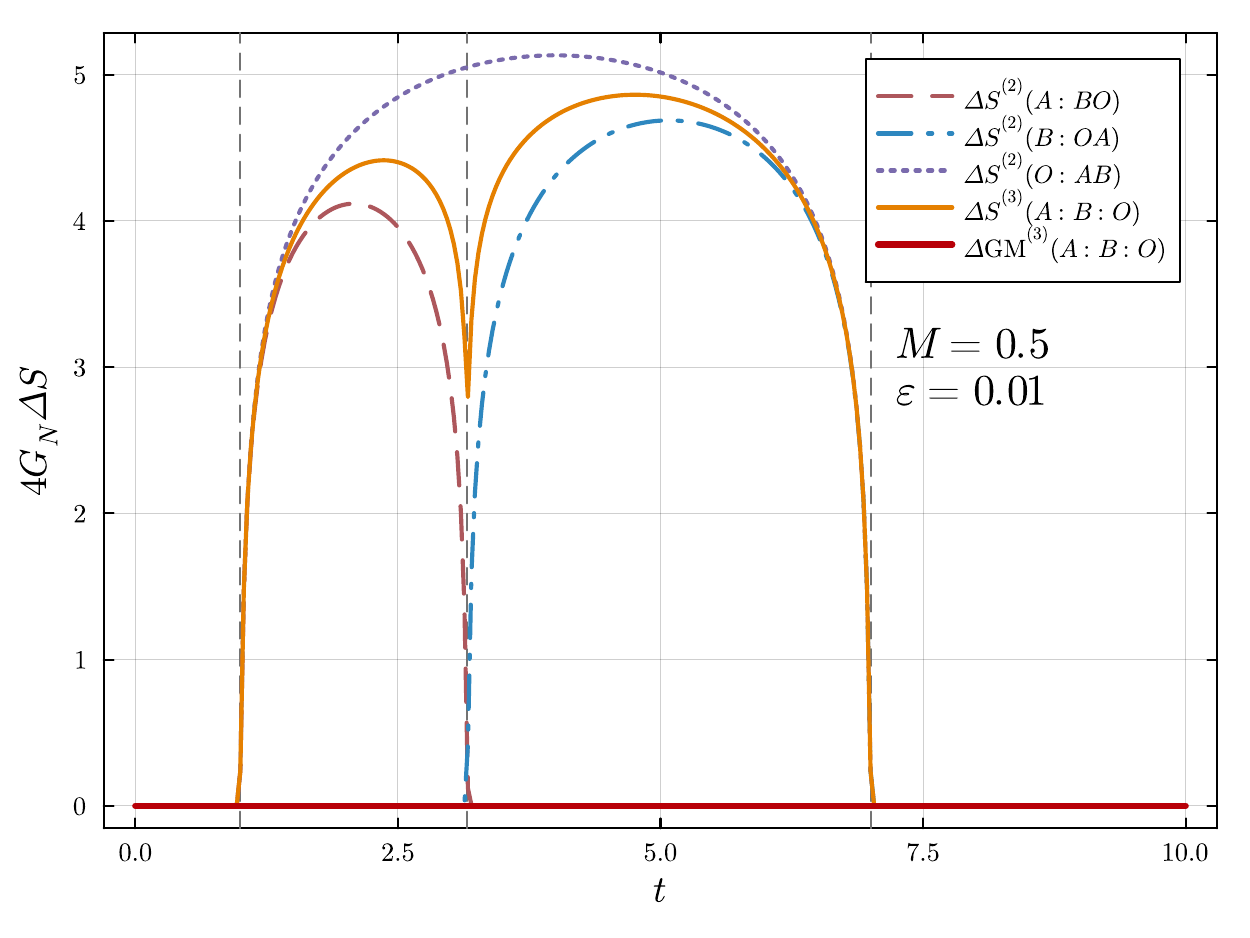}
        \par\smallskip{\small (a) $(x_1,x_2,x_3)=(1,\sqrt{10},7)$\par}
    \end{minipage}
    \hfill
    \begin{minipage}[t]{0.49\textwidth}
        \centering
        \includegraphics[width=\linewidth]{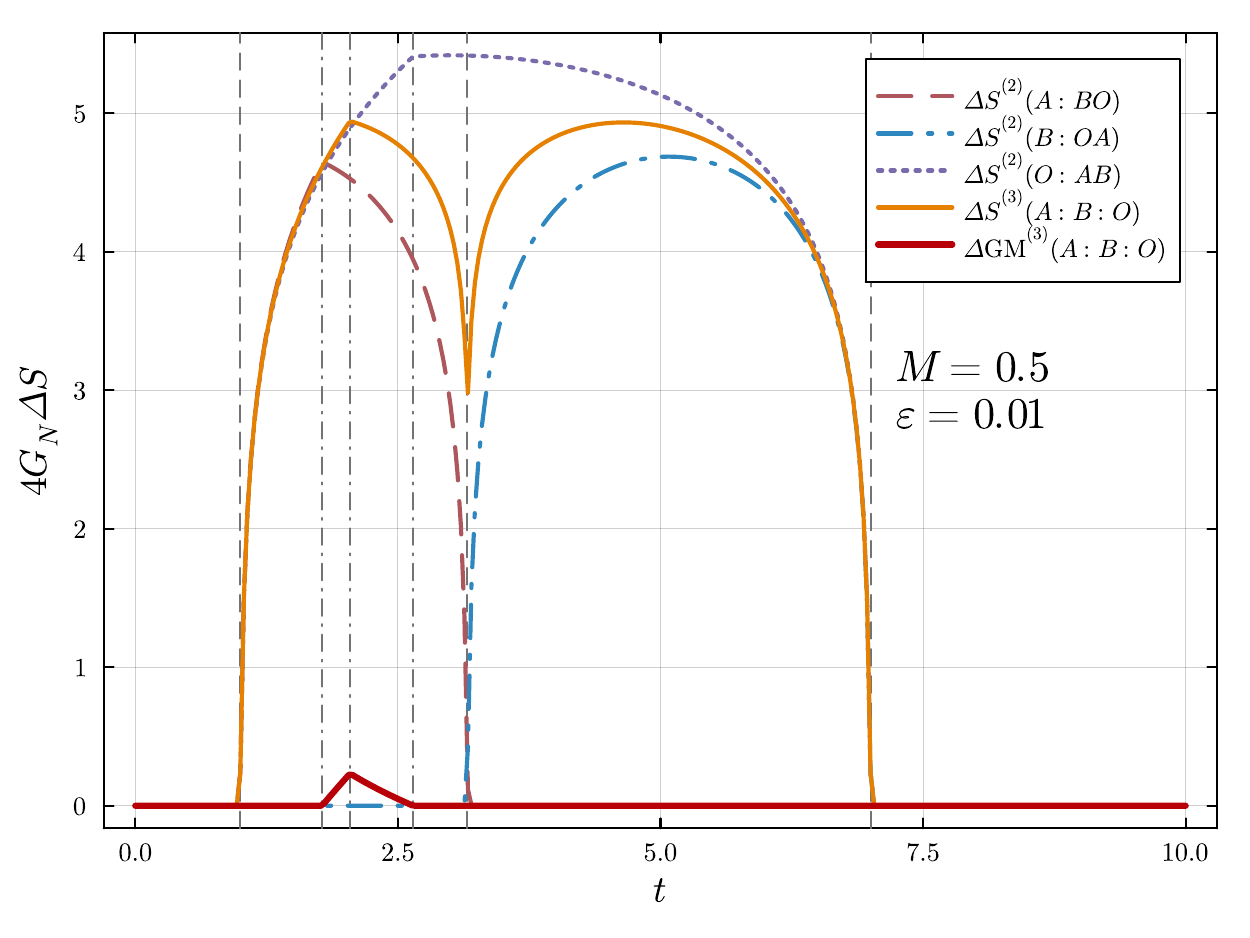}
        \par\smallskip{\small (b) $(x_1,x_2,x_3)=(-1,\sqrt{10},7)$\par}
    \end{minipage}
    \par\medskip
    \begin{minipage}[t]{0.49\textwidth}
        \centering
        \includegraphics[width=\linewidth]{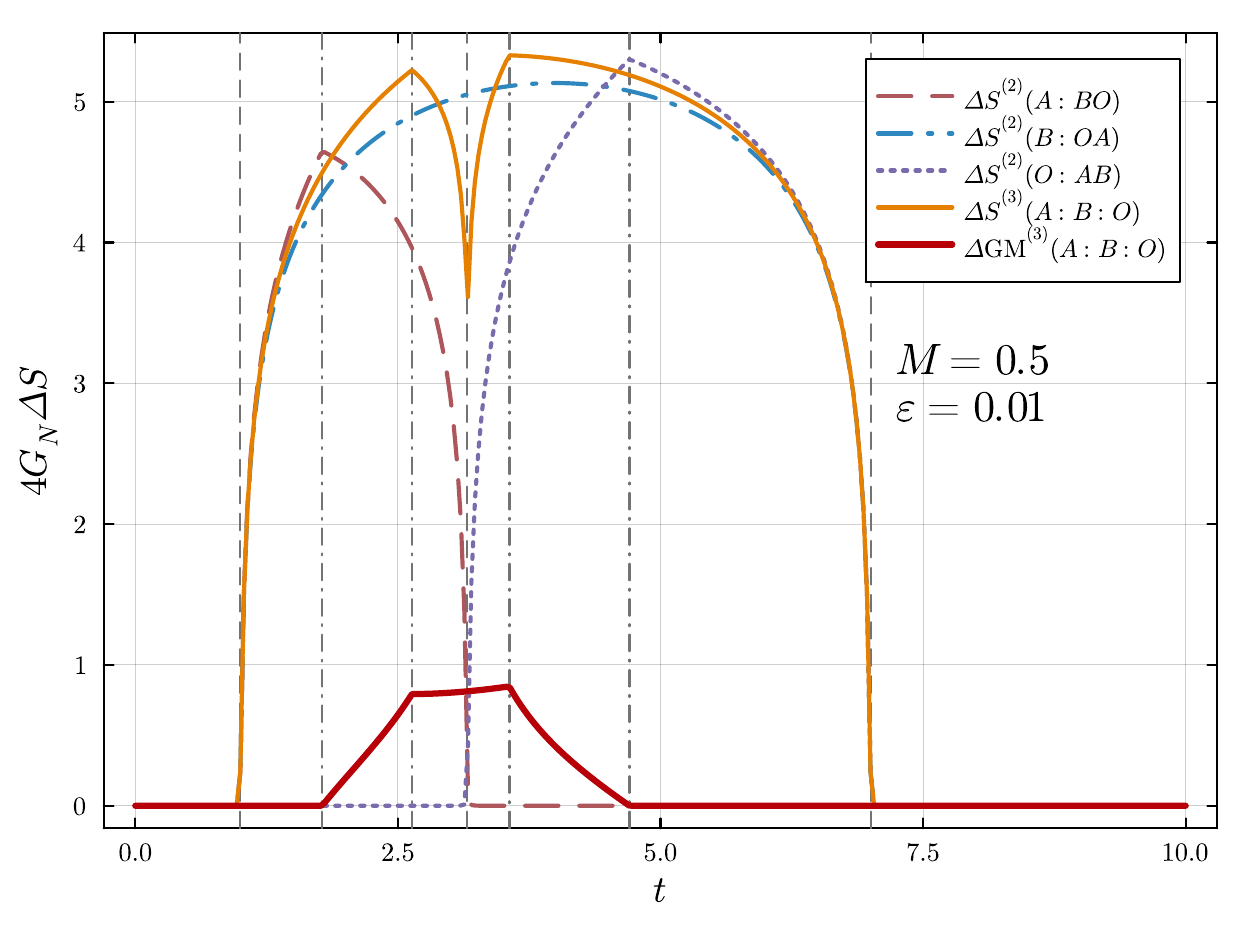}
        \par\smallskip{\small (c) $(x_1,x_2,x_3)=(-\sqrt{10},1,7)$\par}
    \end{minipage}
    \begin{minipage}[t]{0.49\textwidth}
        \centering
        \includegraphics[width=\linewidth]{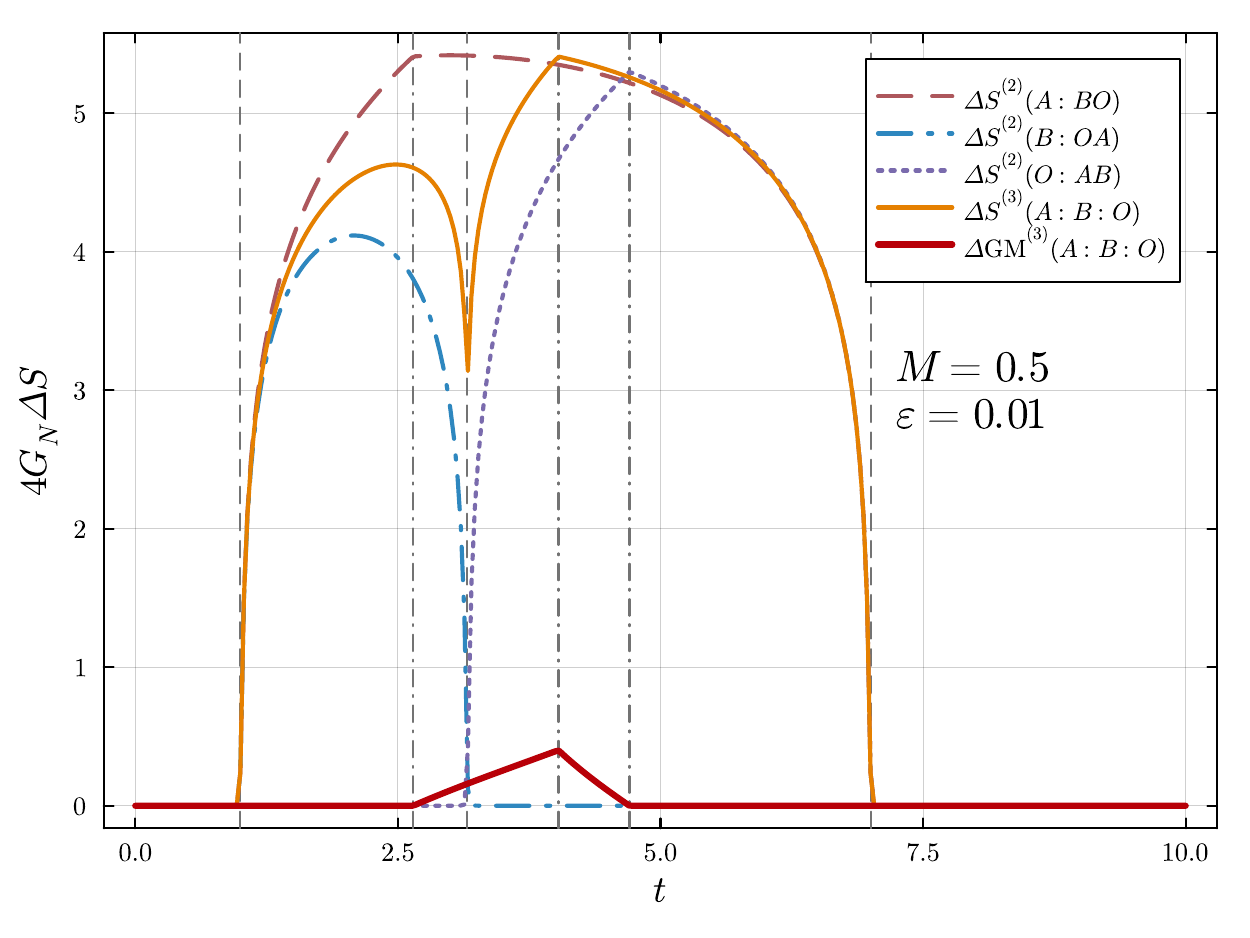}
        \par\smallskip{\small (d) $(x_1,x_2,x_3)=(-7,1,\sqrt{10})$\par}
    \end{minipage}
    \caption{Fully backreacted holographic results in the conical-defect regime for $M=0.5$ and $\varepsilon=0.01$.
    Panels (a)--(d) illustrate cases O, A-I, A-II, and A-III in the main text, respectively.}
    \label{fig_sm:subregionplots_con}
\end{figure}

In applying this formula, we substitute the coordinate transformation into the invariants $\calI_{ij}^{\cd}$.
The time dependence is encoded in the endpoints $(\tau_i, \theta_i, \rho_i)$ and in the corresponding minimizing windings, both of which depend on the physical time $t$.
The numerical results for the conical regime are shown in Fig.~\ref{fig_sm:subregionplots_con}.

\subsection{Winding Transitions and Profile Classification}
Reflection symmetry about the insertion point allows us to take $x_2>0$.
The complete classification consists of the four cases shown in Table~\ref{tab_sm:cases}.
The boundaries are obtained by continuous extension.

\begin{table}[t]
    \centering
    \begin{ruledtabular}
        \begin{tabular}{c c c}
        Case & Endpoint ordering & Trivalent-network sequence\\
        \hline
        O & $0<x_1<x_2<x_3$ & no transition\\
        A-I & $0<-x_1<x_2<x_3$ & $T_A\to T_O$\\
        A-II & $0<x_2<-x_1<x_3$ & $T_A\to T_B\to T_O$\\
        A-III & $0<x_2<x_3<-x_1$ & $T_A\to T_O$
        \end{tabular}
    \end{ruledtabular}
    \caption{Classification of the vacuum-subtracted tripartite GM profiles in the conical-defect regime.}
    \label{tab_sm:cases}
\end{table}

An RT saddle transition changes the minimizing $m_{ij}^{(2)}$, while a trivalent-network transition changes the minimizing class $T$.
The former determines the rise and fall of the vacuum-subtracted tripartite GM, and the latter determines its cusps.

We first determine the RT saddle transitions, which correspond to crossings of the falling particle by the RT geodesics.
In case O, the falling particle is always outside the RT geodesics, so the vacuum-subtracted tripartite GM vanishes at all times.
In case A-I, the change from $m_{12}^{(2)}=0$ to $m_{12}^{(2)}=1$ occurs in the interval $|x_1|<t<x_2$ and obeys
\begin{equation}
    \cos{(\theta_1 - \theta_2)} = \cos{(\theta_1 - \theta_2 + 2\pi\alpha)}.
\end{equation}
With the principal-Arg convention, the condition is
\begin{equation}
    -\pi\alpha = \alpha \left( - \pi + \arctan{\frac{2\varepsilon x_1}{x_1^2 - t^2 - \varepsilon^2}} - \arctan{\frac{2\varepsilon x_2}{x_2^2 - t^2 - \varepsilon^2}}\right).
\end{equation}
Solving this condition gives $t_{12}^{(\varepsilon)} = \sqrt{|x_1|x_2 - \varepsilon^2}$.
The transition lies in the displayed interval when $\varepsilon^2 < |x_1|(x_2 - |x_1|)$, which is satisfied for sufficiently small $\varepsilon$.
Repeating the calculation for the RT geodesics anchored at $x_1$ and $x_3$ gives the transition time $t_{31}^{(\varepsilon)} = \sqrt{|x_1|x_3 - \varepsilon^2}$.
In the sharp limit, these transitions delimit the support window $\sqrt{|x_1|x_2} < t < \sqrt{|x_1|x_3}$ in every nontrivial case.
The standard quasiparticle picture of ballistically propagating entangled pairs does not explain the support of the vacuum-subtracted tripartite GM, which is organized by the winding transitions.

\begin{figure}
    \centering
    \includegraphics[width=0.49\textwidth]{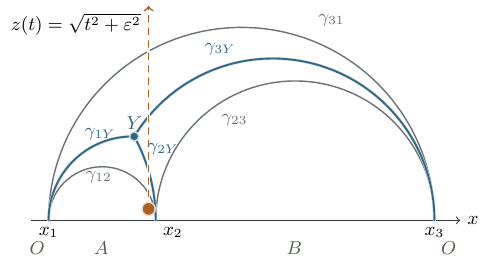}
    \caption{Transition hierarchy in Poincar\'{e} coordinates for case A-II with $0 < x_2 < |x_1| < x_3$.
    Dashed gray curves are RT geodesics and blue curves are trivalent geodesic legs.
    In the sharp-quench limit, the falling particle follows the vertical line and crosses the geodesics in the order $(12)$, $(2Y)$, $(3Y)$, $(31)$.}
    \label{fig_sub:orbits}
\end{figure}

The trivalent branches are the three classes $T_A$, $T_B$, and $T_O$ defined above.
Case O has $T_A$ at all times, case A-I and A-III have the sequence $T_A \to T_O$, and case A-II has $T_A \to T_B \to T_O$.
The representative case A-II is shown in Fig.\ref{fig_sub:orbits}.
The falling particle crosses the RT geodesic for $(12)$, two trivalent geodesic legs $(Y2)$ and $(Y3)$, and the RT geodesic for $(31)$ in that order, which explains the trapezoidal shape of the vacuum-subtracted tripartite GM.

\subsection{Analytic Sharp-quench profiles}
In the sharp-quench limit $\varepsilon \to 0$, the transition points can be determined analytically, and the vacuum-subtracted tripartite GM has an explicit form on each time interval.
In the remainder of this subsection, we assume $x_1 < 0 < x_2 < x_3$.

\paragraph{Endpoint data and windings.}
Away from $t=|x_i|$, the leading sharp-limit expansions are
\begin{align}
    \tilde{\tau}_i(t) &= \begin{cases}
        \displaystyle \frac{2\varepsilon t}{x_i^2 - t^2} + \order{\hat{\varepsilon}^3}, & 0 < t < |x_i|,\\
        \displaystyle \pi - \frac{2\varepsilon t}{t^2 - x_i^2} + \order{\hat{\varepsilon}^3}, & |x_i| < t,
    \end{cases}\label{eq_sm:tauexpansion}\\
    \tilde{\theta}_i(t) &= \sgn(x_i) \begin{cases}
        \displaystyle \frac{2\varepsilon |x_i|}{x_i^2 - t^2} + \order{\hat{\varepsilon}^3}, & 0 < t < |x_i|,\\
        \displaystyle \pi - \frac{2\varepsilon |x_i|}{t^2 - x_i^2} + \order{\hat{\varepsilon}^3}, & |x_i| < t,
    \end{cases}\label{eq_sm:thetaexpansion}
\end{align}
where $\hat{\varepsilon} \coloneqq \varepsilon/\Lambda$ is the dimensionless quench width defined using the smallest length scale $\Lambda$ in the problem, such as $|x_i|$, $t$ or their differences.

For the pair $(12)$, the relevant branches are $m_{12}^{(2)}=0,1$.
The corresponding invariants are then obtained by substituting the expansions \eqref{eq_sm:tauexpansion} and \eqref{eq_sm:thetaexpansion}, yielding
\begin{align}
    \calI_{12}^{\cd}(m_{12}^{(2)}=0) &= \begin{cases}
        \displaystyle 2\alpha^2\varepsilon^2 \frac{(|x_1|+x_2)^2}{(x_1^2-t^2)(x_2^2-t^2)} + \order{\hat{\varepsilon}^4}, & 0 < t < \min(|x_1|,x_2),\\
        \displaystyle 2\alpha \varepsilon \sin{(\pi\alpha)} \frac{|x_1|+x_2}{(t+|x_1|)(x_2-t)} + \order{\hat{\varepsilon}^2}, & |x_1| < t < x_2,\\
        \displaystyle 2\alpha \varepsilon \sin{(\pi\alpha)} \frac{|x_1|+x_2}{(|x_1|-t)(t+x_2)} + \order{\hat{\varepsilon}^2}, & x_2 < t < |x_1|,\\
        \displaystyle 2\sin^2{(\pi\alpha)} + \order{\hat{\varepsilon}}, & \max(|x_1|,x_2) < t,
    \end{cases}\\
    \calI_{12}^{\cd}(m_{12}^{(2)}=1) &= \begin{cases}
        \displaystyle 2\sin^2{(\pi\alpha)} + \order{\hat{\varepsilon}}, & 0 < t < \min(|x_1|,x_2),\\
        \displaystyle 2\alpha \varepsilon \sin{(\pi\alpha)} \frac{|x_1|+x_2}{(t-|x_1|)(t+x_2)} + \order{\hat{\varepsilon}^2}, & |x_1| < t < x_2,\\
        \displaystyle 2\alpha \varepsilon \sin{(\pi\alpha)} \frac{|x_1|+x_2}{(t+|x_1|)(t-x_2)} + \order{\hat{\varepsilon}^2}, & x_2 < t < |x_1|,\\
        \displaystyle 2\alpha^2 \varepsilon^2 \frac{(|x_1|+x_2)^2}{(t^2-x_1^2)(t^2-x_2^2)} + \order{\hat{\varepsilon}^4}, & \max(|x_1|,x_2) < t.
    \end{cases}
\end{align}
For the pair $(23)$, the relevant branches are $m_{23}^{(2)}=0,-1$, and we obtain
\begin{align}
    \calI_{23}^{\cd}(m_{23}^{(2)}=0) &= \begin{cases}
        \displaystyle 2\alpha^2\varepsilon^2 \frac{(x_3-x_2)^2}{(x_2^2-t^2)(x_3^2-t^2)} + \order{\hat{\varepsilon}^4}, & 0 < t < x_2,\\
        \displaystyle 2\alpha \varepsilon \sin{(\pi\alpha)} \frac{x_3-x_2}{(t+x_2)(t+x_3)} + \order{\hat{\varepsilon}^2}, & x_2 < t < x_3,\\
        \displaystyle 2\alpha^2\varepsilon^2 \frac{(x_3-x_2)^2}{(t^2-x_2^2)(t^2-x_3^2)} + \order{\hat{\varepsilon}^4}, & x_3 < t,
    \end{cases}\\
    \calI_{23}^{\cd}(m_{23}^{(2)}=-1) &= \begin{cases}
        \displaystyle 2\sin^2{(\pi\alpha)} + \order{\hat{\varepsilon}}, & 0 < t < x_2,\\
        \displaystyle 2\alpha \varepsilon \sin{(\pi\alpha)} \frac{x_3-x_2}{(t-x_2)(x_3-t)} + \order{\hat{\varepsilon}^2}, & x_2 < t < x_3,\\
        \displaystyle 2\sin^2{(\pi\alpha)} + \order{\hat{\varepsilon}}, & x_3 < t.
    \end{cases}
\end{align}
For the pair $(31)$, the relevant branches are $m_{31}^{(2)}=0,-1$, and we obtain
\begin{align}
    \calI_{31}^{\cd}(m_{31}^{(2)}=0) &= \begin{cases}
        \displaystyle 2\alpha^2\varepsilon^2 \frac{(|x_1|+x_3)^2}{(x_1^2-t^2)(x_3^2-t^2)} + \order{\hat{\varepsilon}^4}, & 0 < t < \min(|x_1|,x_3),\\
        \displaystyle 2\alpha \varepsilon \sin{(\pi\alpha)} \frac{|x_1|+x_3}{(t+|x_1|)(x_3-t)} + \order{\hat{\varepsilon}^2}, & |x_1| < t < x_3,\\
        \displaystyle 2\alpha \varepsilon \sin{(\pi\alpha)} \frac{|x_1|+x_3}{(|x_1|-t)(t+x_3)} + \order{\hat{\varepsilon}^2}, & x_3 < t < |x_1|,\\
        \displaystyle 2\sin^2{(\pi\alpha)} + \order{\hat{\varepsilon}}, & \max(|x_1|,x_3) < t,
    \end{cases}\\
    \calI_{31}^{\cd}(m_{31}^{(2)}=-1) &= \begin{cases}
        \displaystyle 2\sin^2{(\pi\alpha)} + \order{\hat{\varepsilon}}, & 0 < t < \min(|x_1|,x_3),\\
        \displaystyle 2\alpha \varepsilon \sin{(\pi\alpha)} \frac{|x_1|+x_3}{(t-|x_1|)(t+x_3)} + \order{\hat{\varepsilon}^2}, & |x_1| < t < x_3,\\
        \displaystyle 2\alpha \varepsilon \sin{(\pi\alpha)} \frac{|x_1|+x_3}{(t+|x_1|)(t-x_3)} + \order{\hat{\varepsilon}^2}, & x_3 < t < |x_1|,\\
        \displaystyle 2\alpha^2\varepsilon^2 \frac{(|x_1|+x_3)^2}{(t^2-x_1^2)(t^2-x_3^2)} + \order{\hat{\varepsilon}^4}, & \max(|x_1|,x_3) < t.
    \end{cases}
\end{align}

\paragraph{Transitions.}
In the sharp-quench limit $\varepsilon \to 0$, the bipartite minimizations are independent of each other and select the branches as
\begin{align}
    m_{12,\star}^{(2)}(t) &= \begin{cases}
        0, &\quad 0 < t < \sqrt{|x_1|x_2},\\
        1, &\quad \sqrt{|x_1|x_2} < t,
    \end{cases},\\
    m_{23,\star}^{(2)}(t) &= 0,\\
    m_{31,\star}^{(2)}(t) &= \begin{cases}
        0, &\quad 0 < t < \sqrt{|x_1|x_3},\\
        -1, &\quad \sqrt{|x_1|x_3} < t.
    \end{cases}
\end{align}
The three trivalent branches that enter the sharp-quench limit minimization are
\begin{align}
    T_A&=[(0,0,0)], & \bmm^{(3)}(T_A) &= (0,0,0),
    \label{eq:sq-T0}\\
    T_B&=[(0,-1,0)], & \bmm^{(3)}(T_B) &= (1,-1,0),
    \label{eq:sq-T1}\\
    T_O&=[(0,-1,-1)], & \bmm^{(3)}(T_O) &= (1,0,-1).
    \label{eq:sq-T2}
\end{align}
At the isolated transition times, the piecewise expressions below are defined by continuous extension.

\paragraph{Case A-I: $-x_1 < x_2$.}
The trivalent transition is $T_A \to T_O$.
The cusp time $t_<$ is obtained from the condition
\begin{equation}
    \calI_{12}^{\cd}(T_A) \calI_{23}^{\cd}(T_A) \calI_{31}^{\cd}(T_A) = \calI_{12}^{\cd}(T_O) \calI_{23}^{\cd}(T_O) \calI_{31}^{\cd}(T_O),
\end{equation}
and one finds
\begin{equation}
    t_< \coloneq \sqrt{\frac{|x_1|(2x_2x_3 - |x_1|(x_2+x_3))}{x_2+x_3-2|x_1|}}, \qquad \sqrt{|x_1|x_2} < t_< < \sqrt{|x_1|x_3}.
\end{equation}
Substituting the minimizing branches into the formula for the vacuum-subtracted tripartite GM gives
\begin{equation}
    \lim_{\varepsilon \to 0} \varDelta{\gm^{(3)}}(t) = \frac{1}{8G_N} \begin{cases}
        \displaystyle 0, & 0 < t < \sqrt{|x_1|x_2},\\
        \displaystyle \log{\frac{(t-|x_1|)(t+x_2)}{(t+|x_1|)(x_2-t)}}, & \sqrt{|x_1|x_2} < t < t_<,\\
        \displaystyle \log{\frac{(t+|x_1|)(x_3-t)}{(t-|x_1|)(t+x_3)}}, & t_< < t < \sqrt{|x_1|x_3},\\
        \displaystyle 0, & \sqrt{|x_1|x_3} < t.
    \end{cases}
\end{equation}

\paragraph{Case A-II: $x_2 < -x_1 < x_3$.}
The trivalent transitions are $T_A \to T_B \to T_O$.
The first cusp time $t_-$ is obtained from the condition
\begin{equation}
    \calI_{12}^{\cd}(T_A) \calI_{23}^{\cd}(T_A) \calI_{31}^{\cd}(T_A) = \calI_{12}^{\cd}(T_B) \calI_{23}^{\cd}(T_B) \calI_{31}^{\cd}(T_B),
\end{equation}
and one finds
\begin{equation}
    t_- \coloneq \sqrt{\frac{x_2(2|x_1|x_3 - x_2(x_3-|x_1|))}{x_3+2x_2-|x_1|}}, \qquad \sqrt{|x_1|x_2} < t_- < |x_1|.
\end{equation}
The second cusp time $t_+$ is obtained from the condition
\begin{equation}
    \calI_{12}^{\cd}(T_B) \calI_{23}^{\cd}(T_B) \calI_{31}^{\cd}(T_B) = \calI_{12}^{\cd}(T_O) \calI_{23}^{\cd}(T_O) \calI_{31}^{\cd}(T_O),
\end{equation}
and one finds
\begin{equation}
    t_+ \coloneq \sqrt{\frac{x_3(2|x_1|x_2 + x_3(|x_1|-x_2))}{2x_3+x_2-|x_1|}}, \qquad |x_1| < t_+ < \sqrt{|x_1|x_3}.
\end{equation}
Substituting the minimizing branches into the formula for the vacuum-subtracted tripartite GM gives
\begin{equation}
    \lim_{\varepsilon \to 0} \varDelta{\gm^{(3)}}(t) = \frac{1}{8G_N} \begin{cases}
        \displaystyle 0, & 0 < t < \sqrt{|x_1|x_2},\\
        \displaystyle \log{\frac{(t-x_2)(t+|x_1|)}{(|x_1|-t)(t+x_2)}}, & \sqrt{|x_1|x_2} < t < t_-,\\
        \displaystyle \log{\frac{(t+x_2)(t+x_3)}{(t-x_2)(x_3-t)}}, & t_- < t < t_+,\\
        \displaystyle \log{\frac{(t+|x_1|)(x_3-t)}{(t-|x_1|)(t+x_3)}}, & t_+ < t < \sqrt{|x_1|x_3},\\
        \displaystyle 0, & \sqrt{|x_1|x_3} < t.
    \end{cases}
\end{equation}

\paragraph{Case A-III: $x_3 < -x_1$.}
The trivalent transition is $T_A \to T_O$.
The cusp time $t_>$ is obtained from the condition
\begin{equation}
    \calI_{12}^{\cd}(T_A) \calI_{23}^{\cd}(T_A) \calI_{31}^{\cd}(T_A) = \calI_{12}^{\cd}(T_O) \calI_{23}^{\cd}(T_O) \calI_{31}^{\cd}(T_O),
\end{equation}
and one finds
\begin{equation}
    t_> \coloneq \sqrt{\frac{|x_1|(|x_1|(x_2+x_3)-2x_2x_3)}{2|x_1|-x_2-x_3}}, \qquad \sqrt{|x_1|x_2} < t_> < \sqrt{|x_1|x_3}.
\end{equation}
Substituting the minimizing branches into the formula for the vacuum-subtracted tripartite GM gives
\begin{equation}
    \lim_{\varepsilon \to 0} \varDelta{\gm^{(3)}}(t) = \frac{1}{8G_N} \begin{cases}
        \displaystyle 0, & 0 < t < \sqrt{|x_1|x_2},\\
        \displaystyle \log{\frac{(|x_1|+t)(t-x_2)}{(|x_1|-t)(t+x_2)}}, & \sqrt{|x_1|x_2} < t < t_>,\\
        \displaystyle \log{\frac{(|x_1|-t)(t+x_3)}{(t+|x_1|)(t-x_3)}}, & t_> < t < \sqrt{|x_1|x_3},\\
        \displaystyle 0, & \sqrt{|x_1|x_3} < t.
    \end{cases}
\end{equation}

\paragraph{Heavy operator parameter independence.}
The sharp-quench formulas above are logarithms of rational functions of time $t$ and are independent of the chiral weight $h_\psi$.
Their shape is fixed solely by endpoint kinematics and branch ordering.
The analogous expansion in the BTZ regime above the threshold leads to the same log-rational profiles.
This contrasts with the case of a static pure black hole microstate~\cite{Anegawa:2025prn}, where the maximum vacuum-subtracted tripartite GM is set by the Bekenstein--Hawking entropy and therefore retains a nontrivial dependence on the heavy operator parameter.

\subsection{BTZ black hole regime above the threshold}
\begin{figure}
    \centering
    \begin{minipage}[t]{0.49\textwidth}
        \centering
        \includegraphics[width=\linewidth]{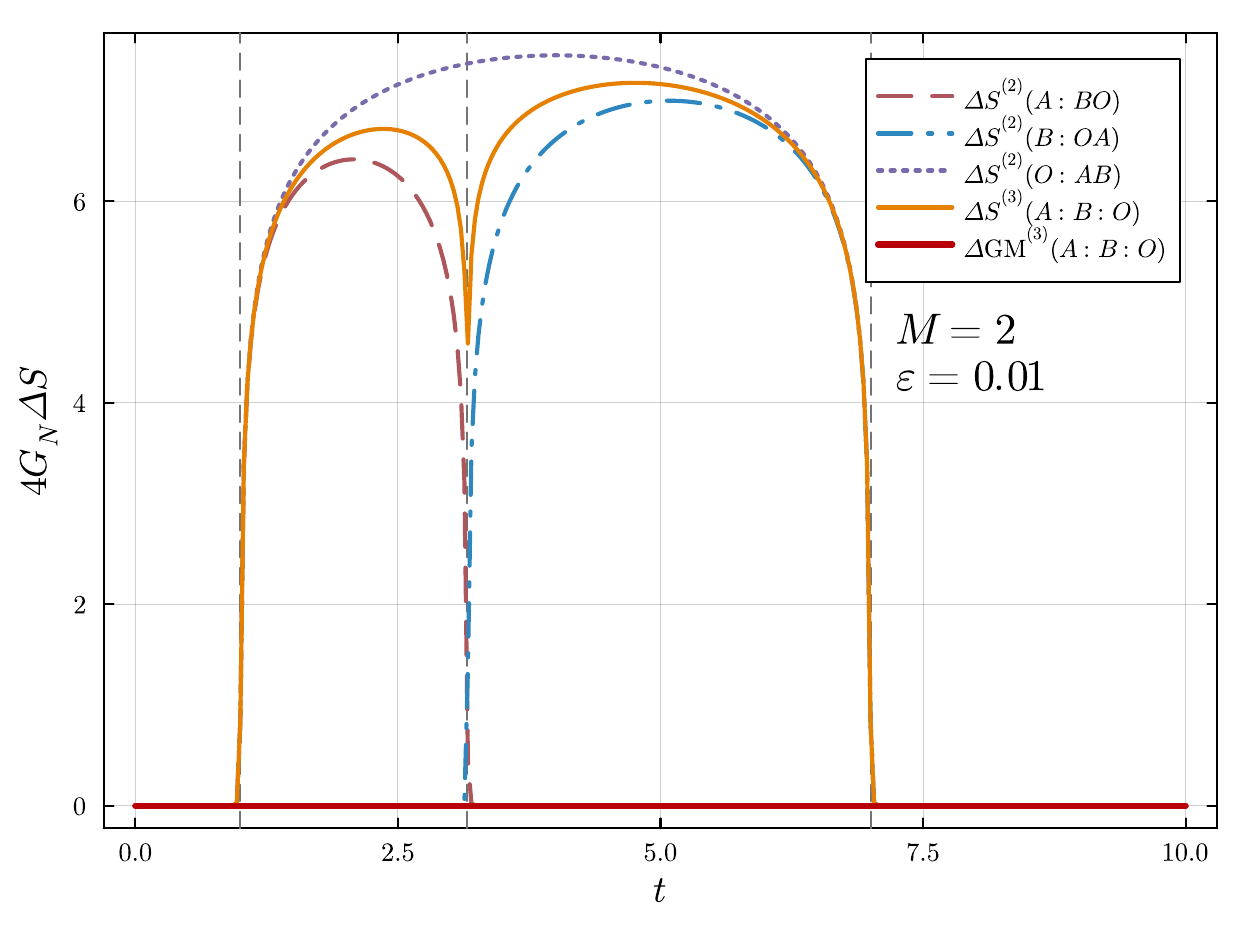}
        \par\smallskip{\small (a) $(x_1,x_2,x_3)=(1,\sqrt{10},7)$\par}
    \end{minipage}
    \hfill
    \begin{minipage}[t]{0.49\textwidth}
        \centering
        \includegraphics[width=\linewidth]{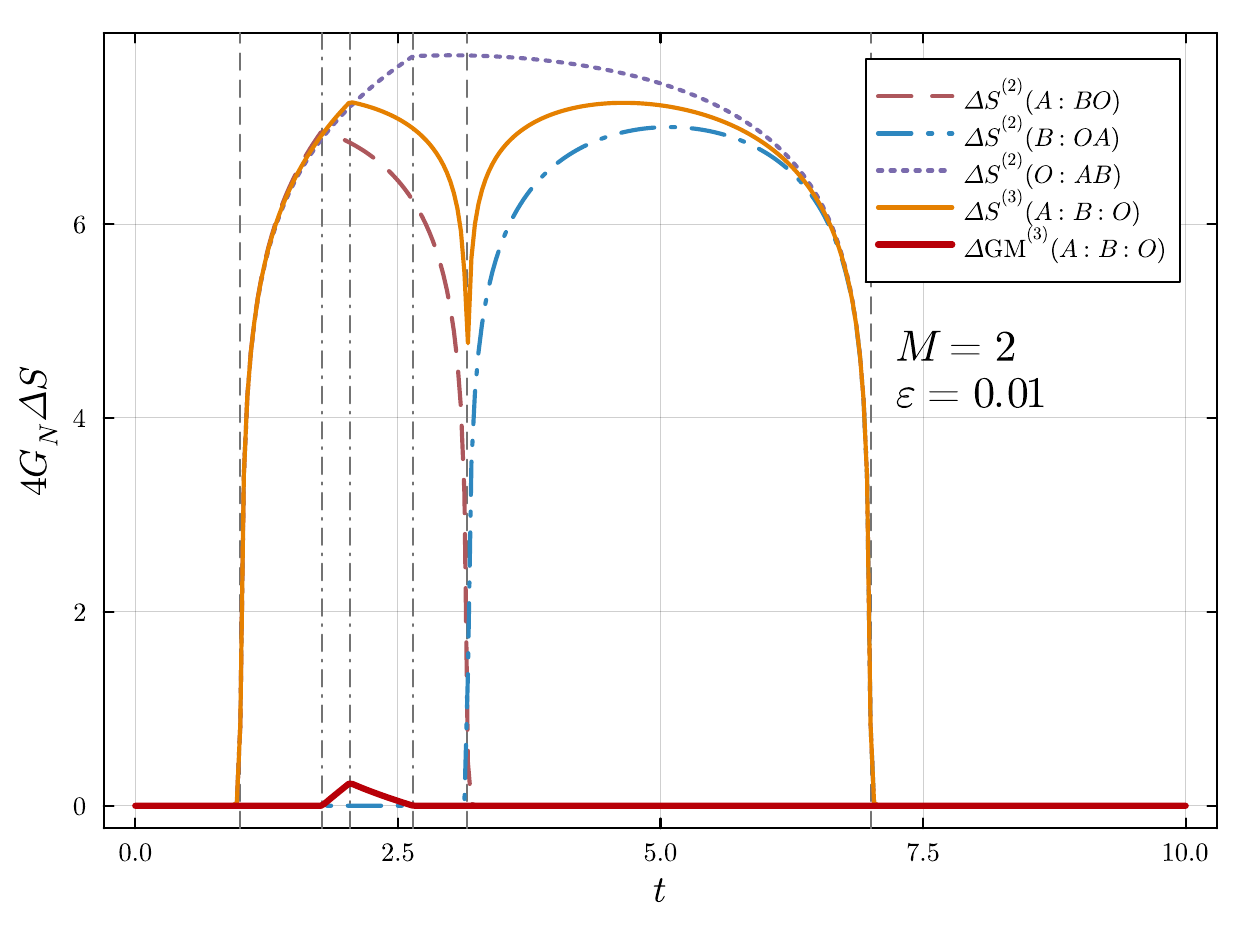}
        \par\smallskip{\small (b) $(x_1,x_2,x_3)=(-1,\sqrt{10},7)$\par}
    \end{minipage}
    \par\medskip
    \begin{minipage}[t]{0.49\textwidth}
        \centering
        \includegraphics[width=\linewidth]{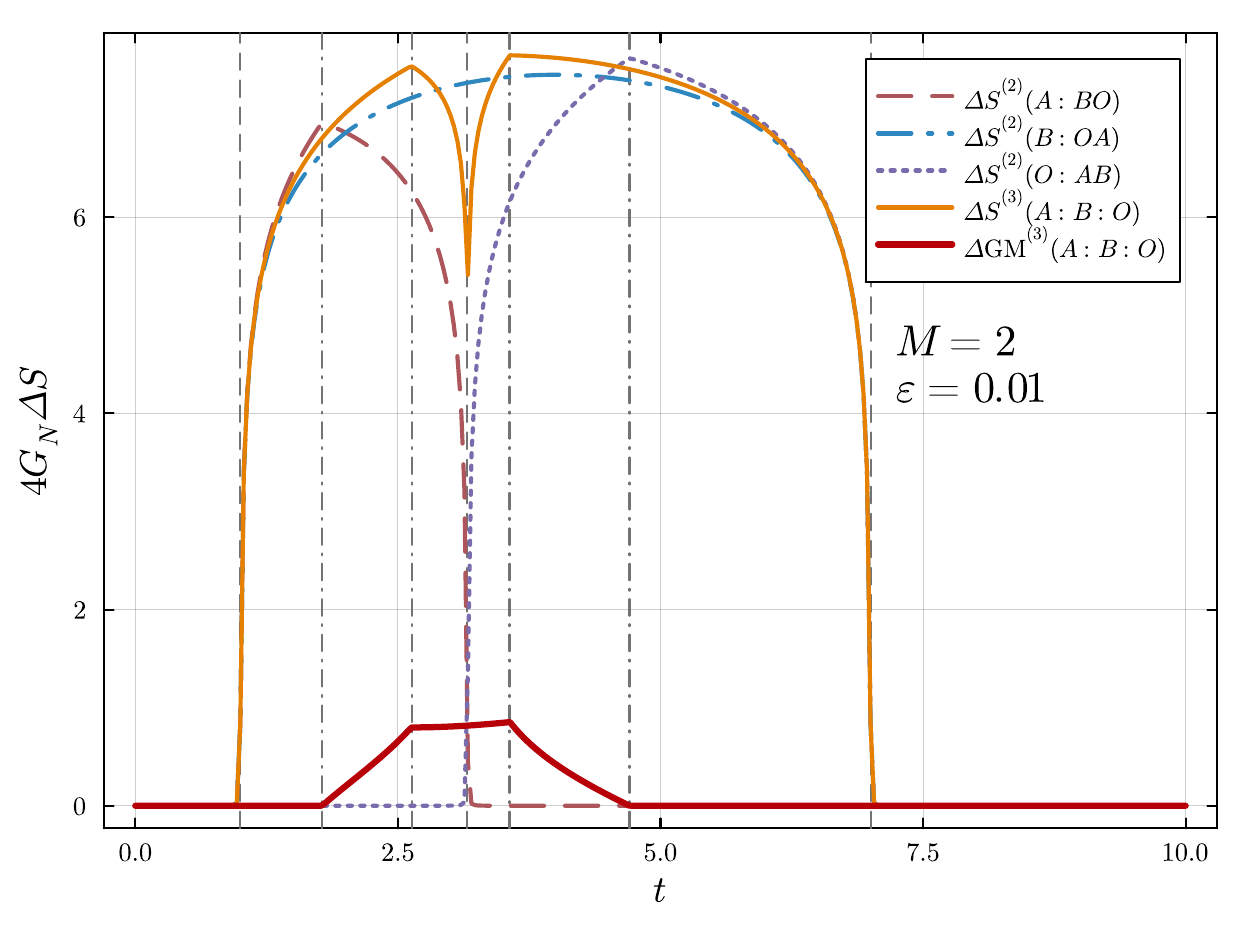}
        \par\smallskip{\small (c) $(x_1,x_2,x_3)=(-\sqrt{10},1,7)$\par}
    \end{minipage}
    \hfill
    \begin{minipage}[t]{0.49\textwidth}
        \centering
        \includegraphics[width=\linewidth]{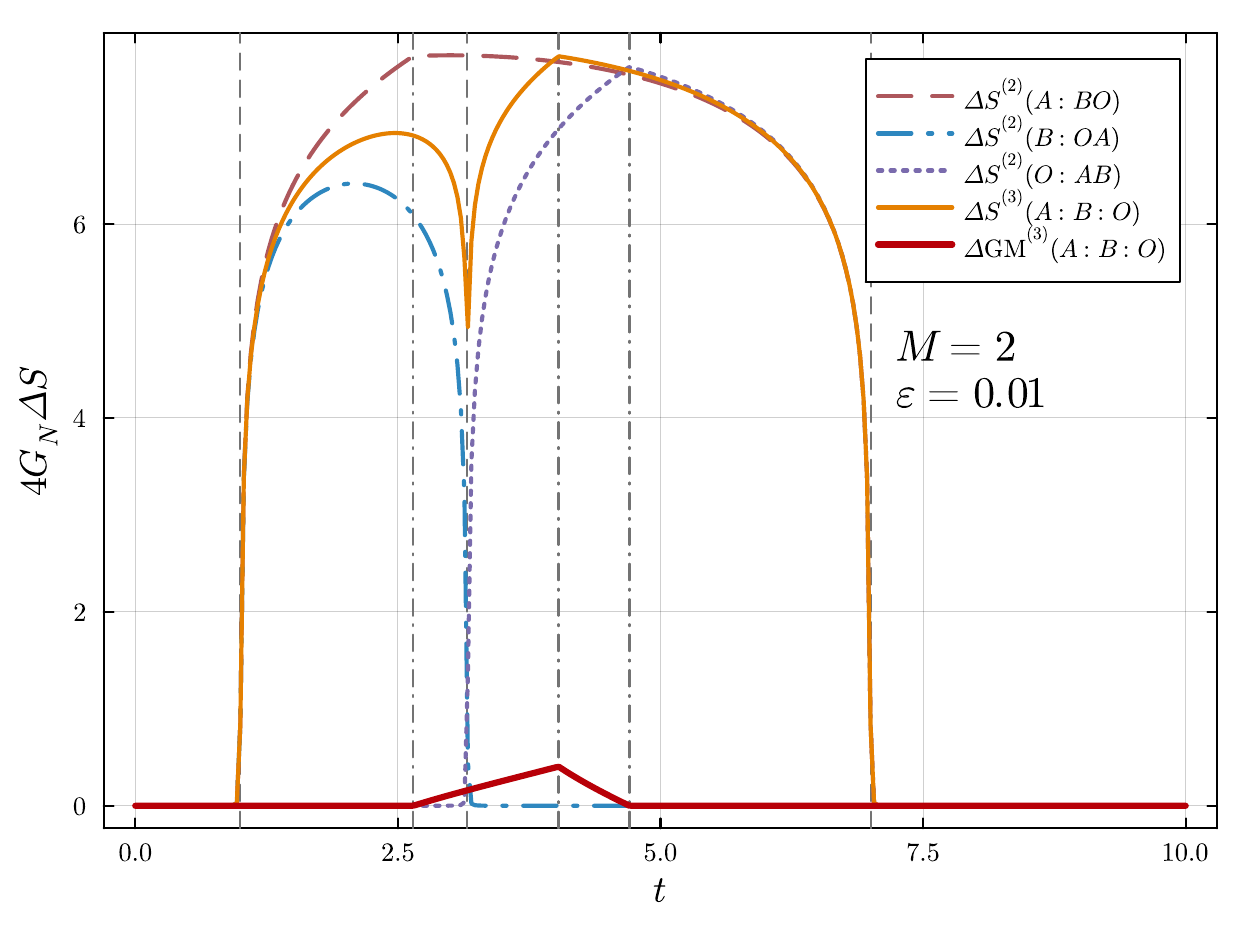}
        \par\smallskip{\small (d) $(x_1,x_2,x_3)=(-7,1,\sqrt{10})$\par}
    \end{minipage}
    \caption{Fully backreacted holographic results in the BTZ regime for $M=2$ and $\varepsilon=0.01$.
    Panels (a)--(d) illustrate cases O, A-I, A-II, and A-III in the main text, respectively.}
    \label{fig_sm:subregionplots_btz}
\end{figure}

For $\alpha = \ii \nu$, define
\begin{equation}
    r = \nu \cosh{\rho}, \qquad \tau = \nu\tilde{\tau}, \qquad \theta = \nu \tilde{\theta}.
\end{equation}
The local metric is
\begin{equation}
    \dd{s}^2 = \dd{\rho}^2 - \sinh^2{\rho} \dd{\tau}^2 + \cosh^2{\rho} \dd{\theta}^2,
\end{equation}
with the periodicity $\theta \sim \theta + 2\pi\nu$.
We use the positive BTZ pair invariant
\begin{equation}
    \calI_{ij}^{\btz}(m_{ij}^{(2)}) = \cosh(\theta_i - \theta_j + 2\pi\nu m_{ij}^{(2)}) - \cosh(\tau_i - \tau_j).
\end{equation}
The trivalent network can be analyzed analogously to the conical-defect regime, using
\begin{equation}
    \calI_{ij}^{\btz}(T) \coloneqq \calI_{ij}^{\btz}(m_{ij}^{(3)}(T)) = \cosh(\vartheta_i-\vartheta_j)-\cosh(\tau_i-\tau_j).
\end{equation}
The same stationarity equations apply with $\calI_{ij}^{\cd}$ replaced by $\calI_{ij}^{\btz}$.
The vacuum-subtracted tripartite GM is given by Eq.~\eqref{eq_sm:bulk-excess-gme} with the BTZ invariants and domains.
The numerical results for the BTZ regime are shown in Fig.~\ref{fig_sm:subregionplots_btz}.

\subsection{Finite-width and mass dependence}
\begin{figure}
    \centering
    \begin{minipage}{0.49\textwidth}
        \centering
        \includegraphics[width=\linewidth]{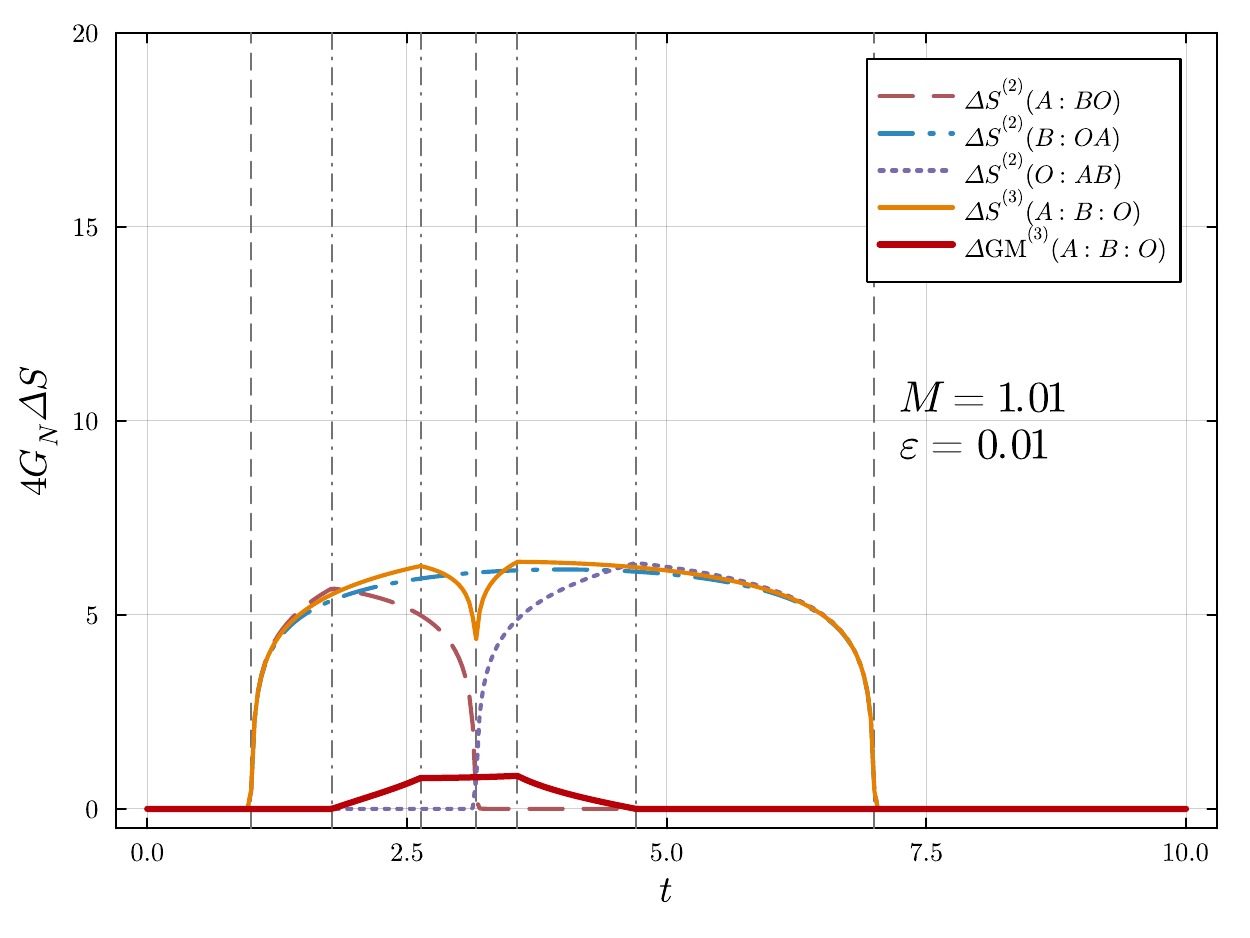}
        \par\smallskip{\small (a) $M=1.01$}
    \end{minipage}
    \hfill
    \begin{minipage}{0.49\textwidth}
        \centering
        \includegraphics[width=\linewidth]{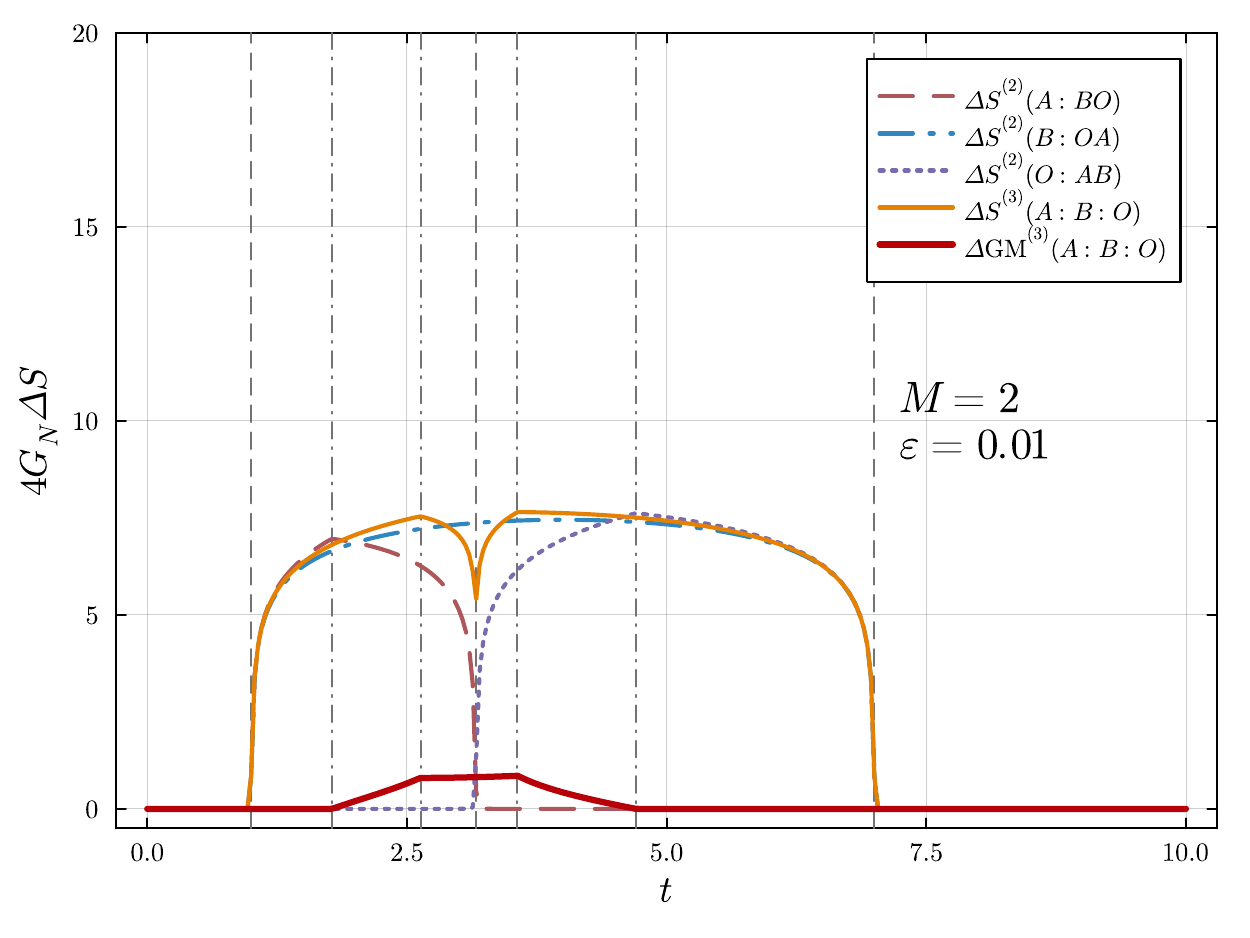}
        \par\smallskip{\small (b) $M=2$}
    \end{minipage}
    \par\medskip
    \begin{minipage}{0.49\textwidth}
        \centering
        \includegraphics[width=\linewidth]{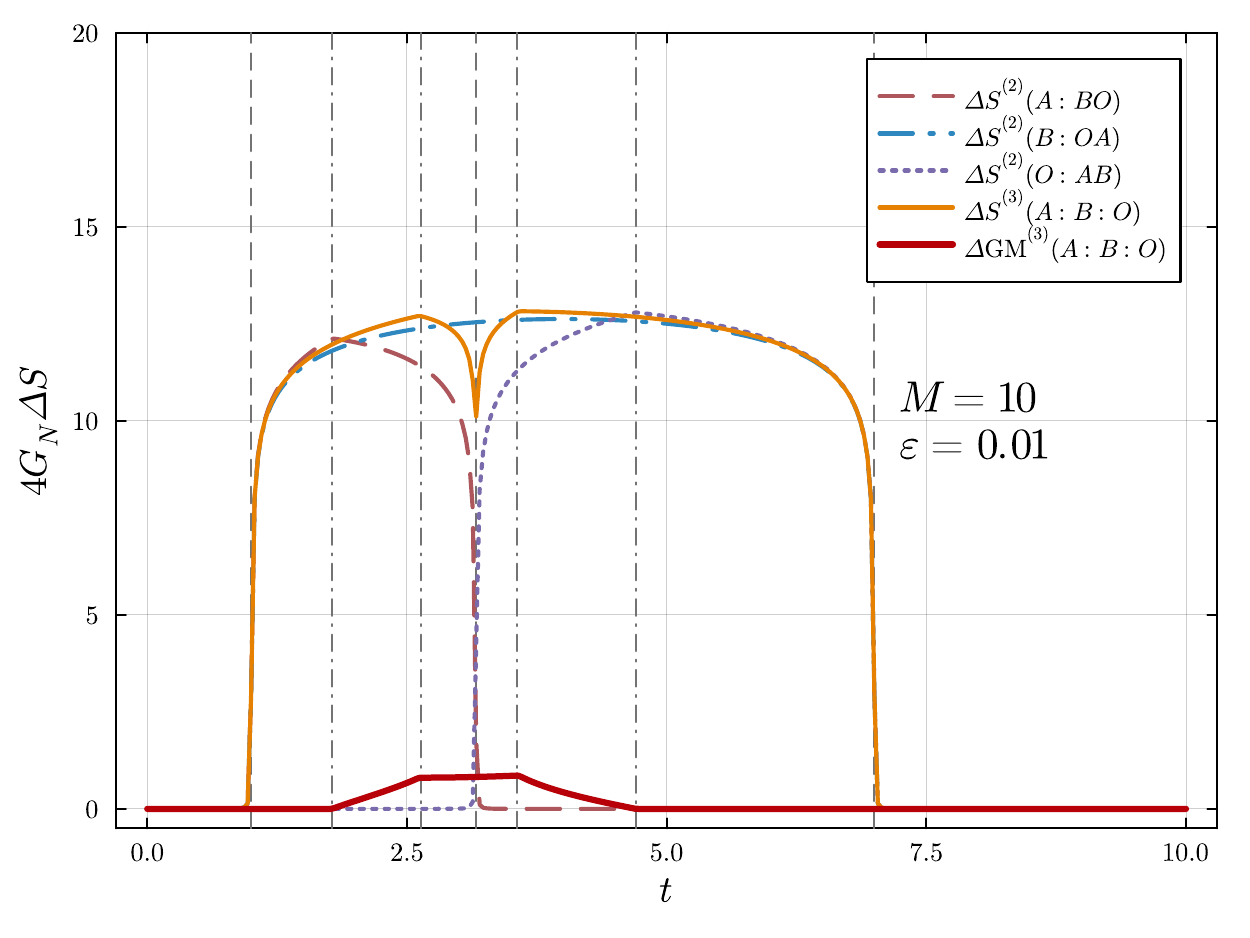}
        \par\smallskip{\small (c) $M=10$}
    \end{minipage}
    \hfill
    \begin{minipage}{0.49\textwidth}
        \centering
        \includegraphics[width=\linewidth]{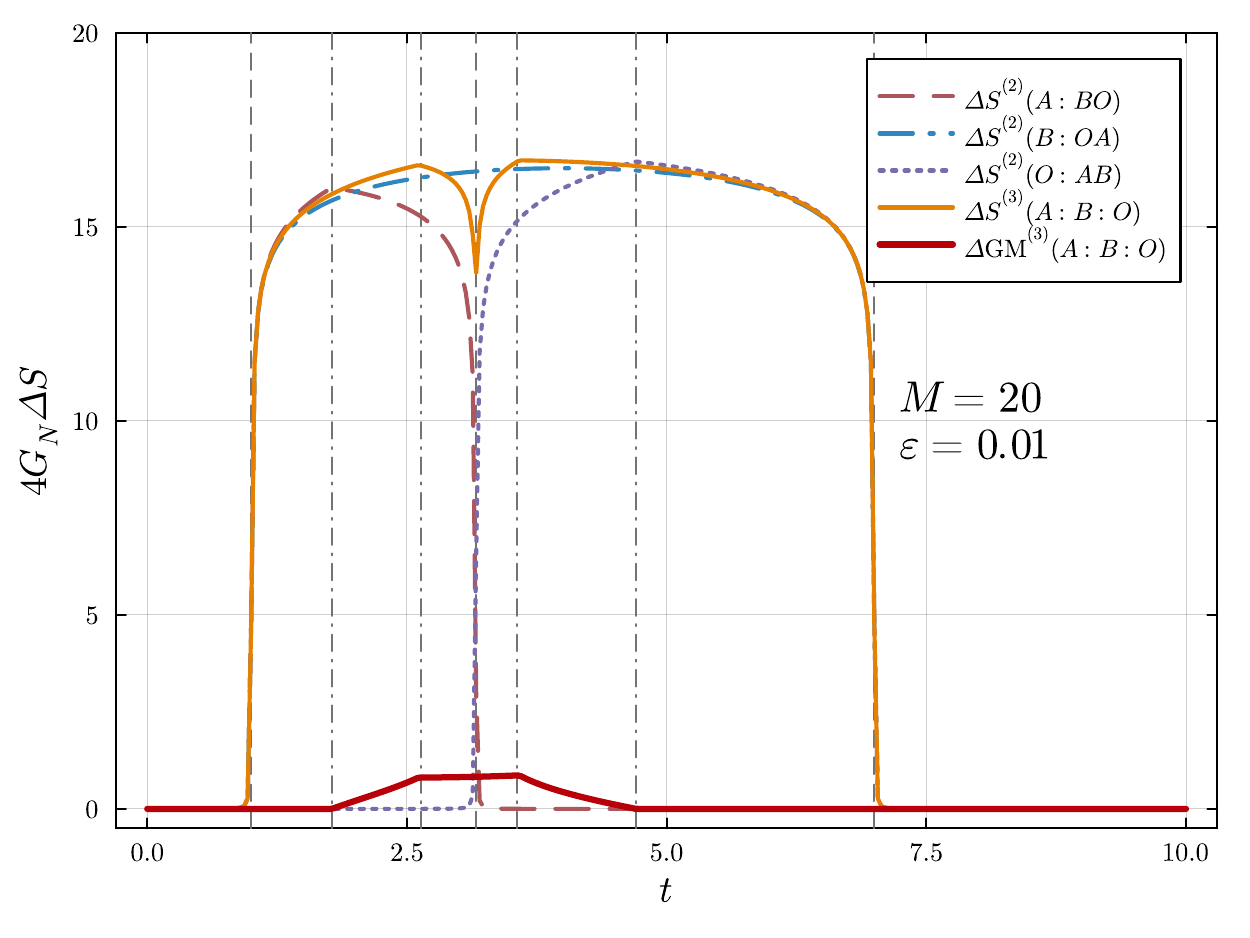}
        \par\smallskip{\small (d) $M=20$}
    \end{minipage}
    \caption{Fully back-reacted holographic entropies for different mass parameters in the BTZ regime. We set $(x_1,x_2,x_3)=(-\sqrt{10},1,7)$, and $\varepsilon=0.01$.}
    \label{fig_sm:Mplots_paper}
\end{figure}

Fig.~\ref{fig_sm:Mplots_paper} shows scans over the mass parameter.
The vacuum-subtracted tripartite GM changes only mildly while the raw entropies increase, and the displayed transition times are insensitive to the mass.

\begin{figure}
    \centering
    \begin{minipage}{0.49\textwidth}
        \centering
        \includegraphics[width=\linewidth]{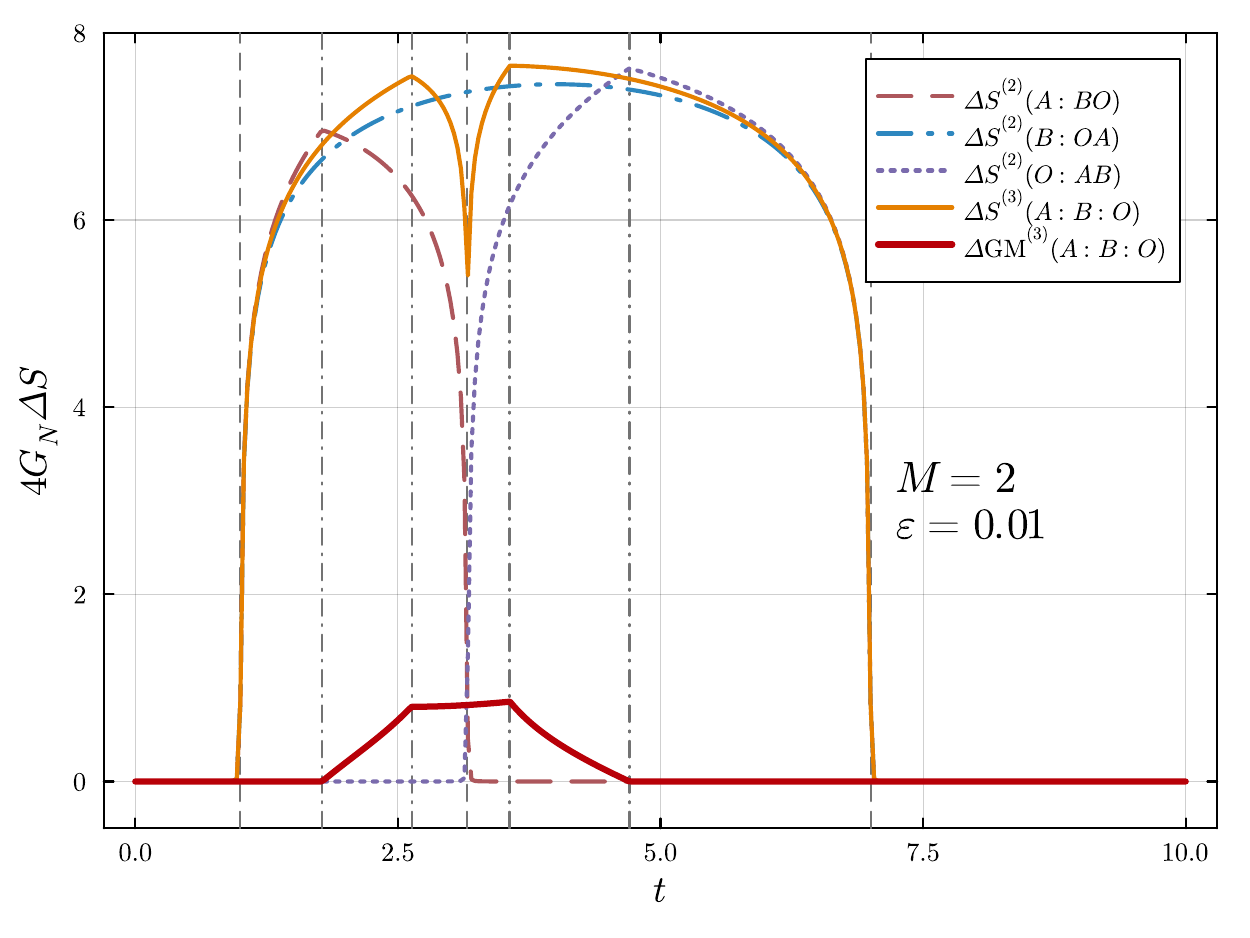}
        \par\smallskip{\small (a) $\varepsilon=0.01$}
    \end{minipage}
    \hfill
    \begin{minipage}{0.49\textwidth}
        \centering
        \includegraphics[width=\linewidth]{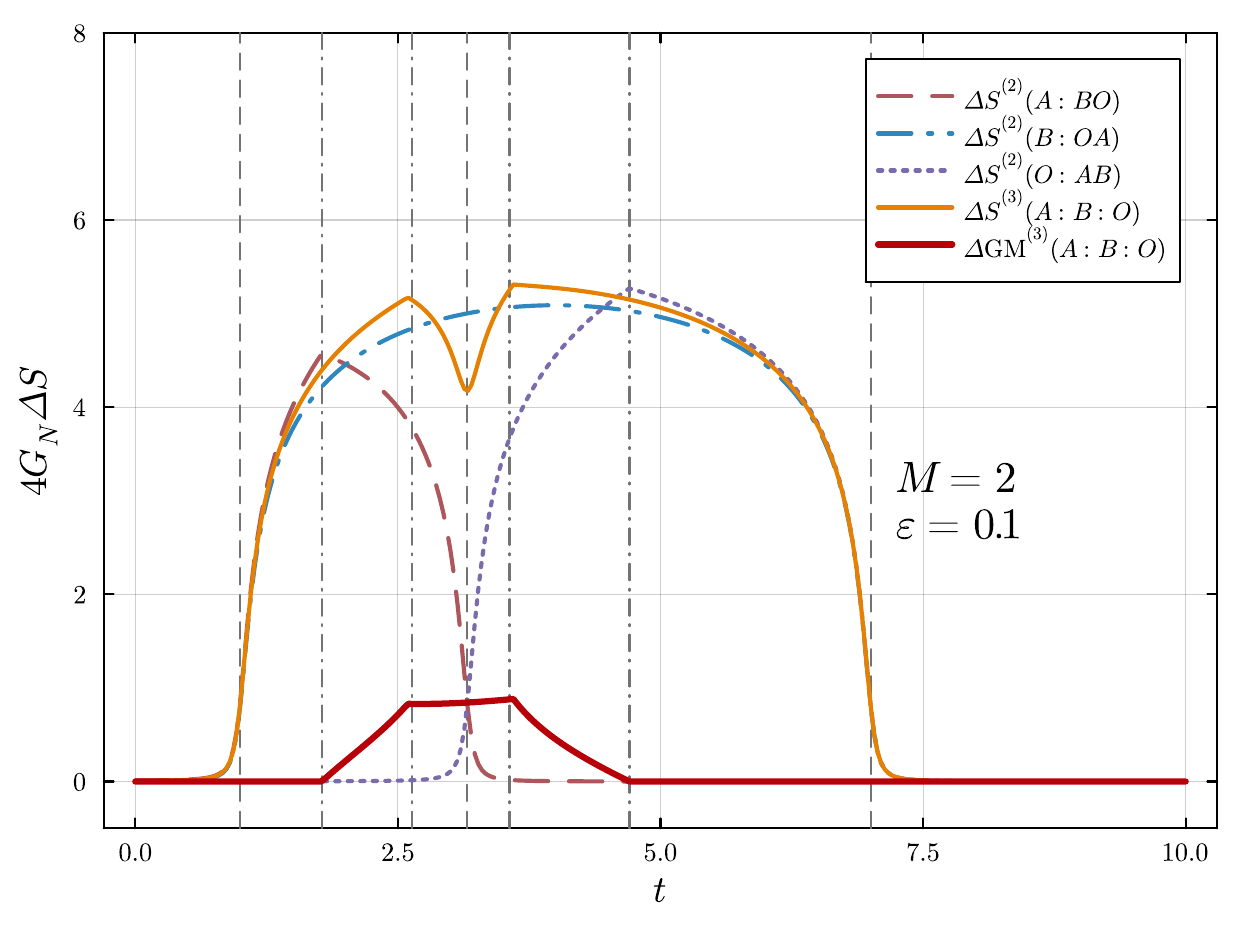}
        \par\smallskip{\small (b) $\varepsilon=0.1$}
    \end{minipage}
    \par\medskip
    \begin{minipage}{0.49\textwidth}
        \centering
        \includegraphics[width=\linewidth]{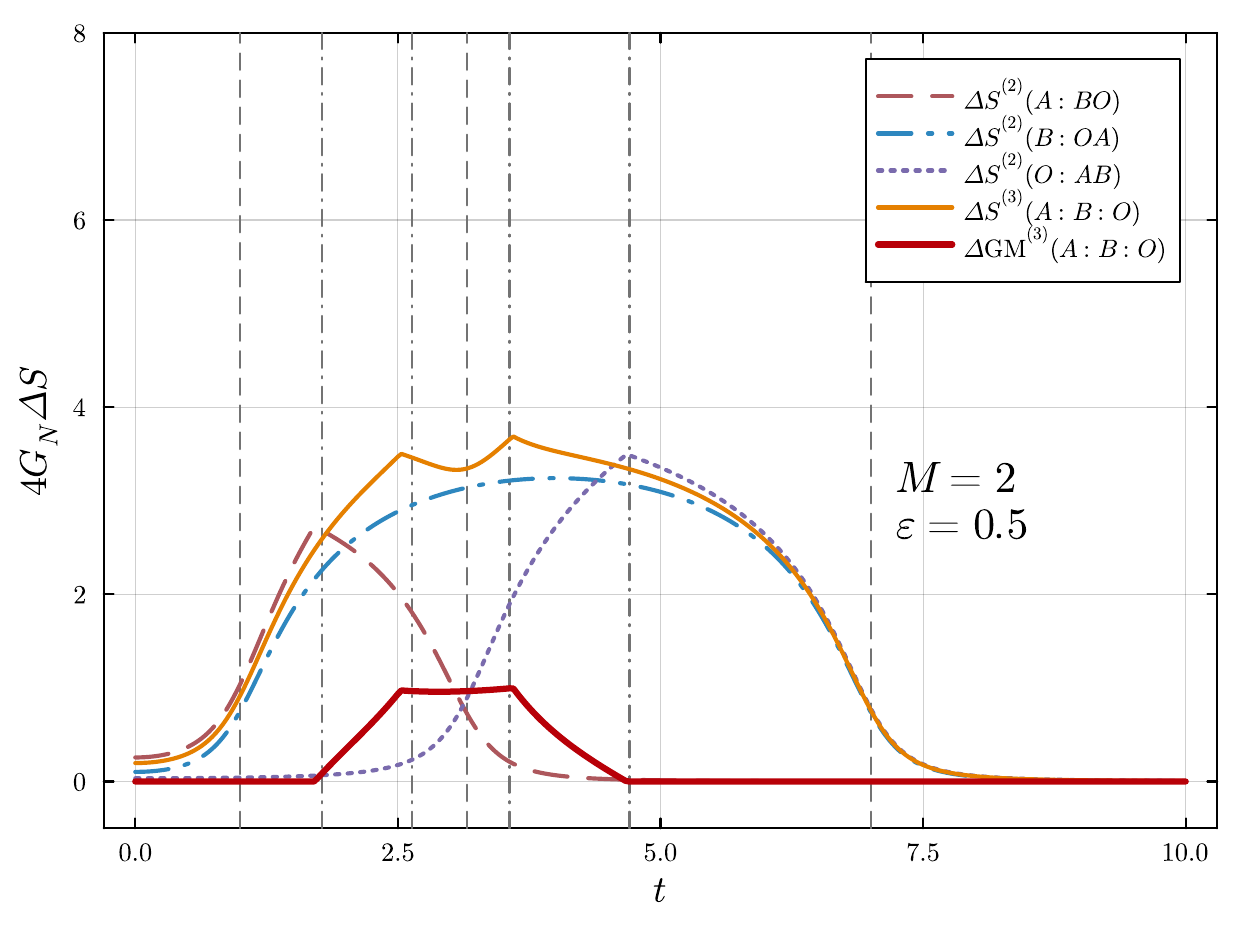}
        \par\smallskip{\small (c) $\varepsilon=0.5$}
    \end{minipage}
    \hfill
    \begin{minipage}{0.49\textwidth}
        \centering
        \includegraphics[width=\linewidth]{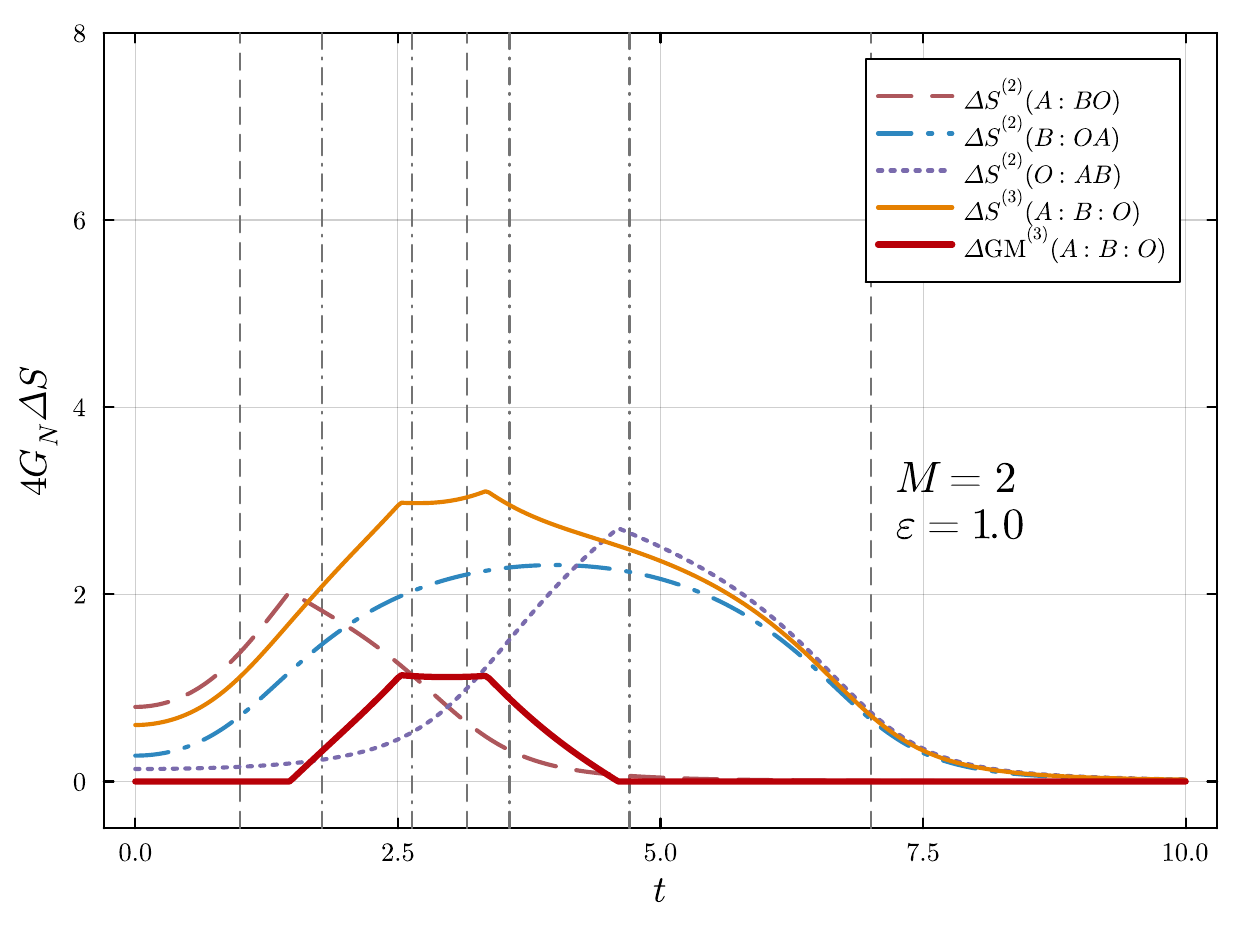}
        \par\smallskip{\small (d) $\varepsilon=1$}
    \end{minipage}
    \caption{Fully back-reacted holographic entropies for different quench widths in the BTZ regime. We set $(x_1,x_2,x_3)=(-\sqrt{10},1,7)$, and $M=2$.}
    \label{fig_sm:deltaplots_paper}
\end{figure}

Fig.~\ref{fig_sm:deltaplots_paper} shows scans over the quench width.
The vacuum-subtracted tripartite GM changes mildly compared with the raw entropies, and its onset sharpens as quench width decreases, consistent with the interpretation of the width as the size of the local excitation~\cite{Nozaki:2013wia}.

\section{Conformal Field Theory Analysis}
We now turn to the CFT analysis of the entropies.
The heavy-light block gives the same branch-ratio formula~\eqref{eq_sm:bulk-excess-gme}, with CFT branch choices identified with the bulk winding sectors.

The calculation is performed in a large-$c$ CFT with a sparse light spectrum and fixed ratio $h_\psi/c$, with twist operators treated as light near the $n\to1$ limit.
We assume vacuum Virasoro block dominance, branch-independent twist OPE coefficients, and diagonal real Lorentzian branches.

\subsection{Entropies from heavy-light approximation}
\paragraph{Bipartite entanglement entropy.}
For a single interval with endpoints $x_i$ and $x_j$, define the normalized correlator
\begin{equation}
    Z_n^{(2)}(x_i, x_j) \coloneqq \frac{\expval{(\Psi_n^{(2)})^\dagger(w_\infty) \calT_{g_A^{(2)}}(w_i) \calT_{(g_A^{(2)})^{-1}}(w_j) \Psi_n^{(2)}(w_0)}}{\expval{(\Psi_n^{(2)})^\dagger(w_\infty) \Psi_n^{(2)}(w_0)}}.
\end{equation}
Choose the M\"obius map
\begin{equation}
    w \mapsto \zeta(w) \coloneqq \frac{w-w_0}{w-w_\infty},
    \label{eq_sm:mobius-map}
\end{equation}
so that the operator insertions are mapped to
\begin{equation}
    w_0 \mapsto 0, \quad w_\infty \mapsto \infty.
\end{equation}
Conformal covariance gives
\begin{equation}
    Z_n^{(2)}(x_i, x_j) = \left| \frac{\zeta_{ij}^2}{w_{ij}^2} \right|^{2h_n^{(2)}} \frac{\expval{(\Psi_n^{(2)})^\dagger(\infty) \calT_{g_A^{(2)}}(\zeta_i) \calT_{(g_A^{(2)})^{-1}}(\zeta_j) \Psi_n^{(2)}(0)}}{\expval{(\Psi_n^{(2)})^\dagger(\infty) \Psi_n^{(2)}(0)}},
\end{equation}
where we used
\begin{equation}
    \eval{\dv{\zeta}{w}}_{w=w_i} \eval{\dv{\zeta}{w}}_{w=w_j} = \frac{\zeta_{ij}^2}{w_{ij}^2}.
\end{equation}

To leading order in $(n-1)$, the heavy-light vacuum block gives~\cite{Asplund:2014coa,Fitzpatrick:2014vua,Fitzpatrick:2015zha,Perlmutter:2015iya,Alkalaev:2015fbw}
\begin{equation}
    \frac{\expval{(\Psi_n^{(2)})^\dagger (\infty) \calT_{g_A^{(2)}}(\zeta_i) \calT_{(g_A^{(2)})^{-1}}(\zeta_j) \Psi_n^{(2)}(0)}}{\expval{(\Psi_n^{(2)})^\dagger(\infty) \Psi_n^{(2)}(0)}} \simeq C_n^{(2)} \left| \frac{u_i' u_j'}{u_{ij}^2} \right|^{2h_n^{(2)}},
\end{equation}
where $u(z) \coloneqq \zeta^\alpha$ is the uniformization map, and $C_n^{(2)}$ is the branch-independent two-point twist normalization.
In total, we obtain
\begin{equation}
    Z_n^{(2)}(x_i, x_j) \simeq C_n^{(2)} \left| \frac{\zeta_{ij}^2 u_i' u_j'}{w_{ij}^2 u_{ij}^2} \right|^{2h_n^{(2)}}.
\end{equation}
Subtracting the vacuum value gives the vacuum-subtracted bipartite entanglement entropy
\begin{equation}
    \varDelta{S}^{(2)}(i,j) = \frac{c}{6} \log{\left| \frac{u_{ij}^2}{\zeta_{ij}^2 u_i' u_j'} \right|}.
    \label{eq_sm:cft-bi-excess}
\end{equation}
This formula is branch-dependent because the uniformization map is multi-valued.

\paragraph{Tripartite ME.}
The five-point function for the tripartite ME is
\begin{equation}
    Z_n^{(3)}(x_1, x_2, x_3) \coloneqq \frac{\expval{(\Psi_n^{(3)})^\dagger(w_\infty) \calT_{g_A}(w_1) \calT_{g_A^{-1}g_B}(w_2) \calT_{g_B^{-1}}(w_3) \Psi_n^{(3)}(w_0)}}{\expval{(\Psi_n^{(3)})^\dagger(w_\infty) \Psi_n^{(3)}(w_0)}}.
\end{equation}
The M\"obius map \eqref{eq_sm:mobius-map} maps the five-point function to
\begin{equation}
    Z_n^{(3)}(x_1, x_2, x_3) = \left| \frac{\zeta_{12} \zeta_{23} \zeta_{31}}{w_{12} w_{23} w_{31}} \right|^{2h_n^{(3)}} \frac{\expval{(\Psi_n^{(3)})^\dagger (\infty) \calT_{g_A}(\zeta_1) \calT_{g_A^{-1}g_B}(\zeta_2) \calT_{g_B^{-1}}(\zeta_3) \Psi_n^{(3)}(0)}}{\expval{(\Psi_n^{(3)})^\dagger(\infty) \Psi_n^{(3)}(0)}},
\end{equation}
where we used M\"obius covariance of the three-point function,
\begin{equation}
    \eval{\dv{\zeta}{w}}_{w=w_1} \eval{\dv{\zeta}{w}}_{w=w_2} \eval{\dv{\zeta}{w}}_{w=w_3} = \frac{\zeta_{12} \zeta_{23} \zeta_{31}}{w_{12} w_{23} w_{31}}.
\end{equation}
Approximating the higher-point heavy state correlator by the heavy-light vacuum block using the uniformization map, we obtain
\begin{equation}
    \frac{\expval{(\Psi_n^{(3)})^\dagger (\infty) \calT_{g_A}(\zeta_1) \calT_{g_A^{-1}g_B}(\zeta_2) \calT_{g_B^{-1}}(\zeta_3) \Psi_n^{(3)}(0)}}{\expval{(\Psi_n^{(3)})^\dagger(\infty) \Psi_n^{(3)}(0)}} \simeq C_n^{(3)} \left| \frac{u_1' u_2' u_3'}{u_{12} u_{23} u_{31}} \right|^{2h_n^{(3)}}.
\end{equation}
Subtracting the vacuum value gives the vacuum-subtracted tripartite ME
\begin{equation}
    \varDelta{S}^{(3)} = \frac{c}{6} \log{\left| \frac{u_{12} u_{23} u_{31}}{\zeta_{12} \zeta_{23} \zeta_{31} u_1' u_2' u_3'} \right|}.
    \label{eq_sm:cft-tri-excess}
\end{equation}
The OPE coefficient $C_n^{(3)}$ is the same branch-independent twist OPE coefficient that appears in the vacuum three-point function and therefore cancels in the vacuum-subtracted tripartite ME.
This formula is also branch-dependent.

\subsection{Branch phases and saddle matching}
After analytic continuation to Lorentzian time, $w_i=x_i-t$ and $\bar{w}_i=x_i+t$ for endpoint insertions at time $t$.
The M\"obius map then maps the endpoint twist insertions to points on the unit circle in the $\zeta$-plane with phases $\phi_i$ and $\bar{\phi}_i$ defined by
\begin{equation}
    \zeta_i = \frac{w_i + \ii \varepsilon}{w_i - \ii \varepsilon} = e^{\ii \phi_i}, \qquad \bar{\zeta}_i = \frac{\bar{w}_i - \ii \varepsilon}{\bar{w}_i + \ii \varepsilon} = e^{-\ii \bar{\phi}_i}.
\end{equation}
The multi-valued uniformization is described by lifted phases
\begin{equation}
    \varphi_i = \phi_i + 2\pi n_i, \qquad \bar{\varphi}_i = \bar{\phi}_i + 2\pi \bar{n}_i, \qquad n_i, \bar{n}_i \in \bbZ,
\end{equation}
so that the uniformization map is single-valued in terms of $\varphi_i$ and $\bar{\varphi}_i$,
\begin{equation}
    u_i = e^{\ii \alpha \varphi_i}, \qquad \bar{u}_i = e^{-\ii \alpha \bar{\varphi}_i}.
\end{equation}
Using the coordinate transformation \eqref{eq_sm:poincare-global-map}, the diagonal branch choice gives the relation between the lifted phases and the bulk angular coordinates as
\begin{equation}
    \alpha \varphi_i = \vartheta_i + \tau_i, \qquad \alpha \bar{\varphi}_i = \vartheta_i - \tau_i,
\end{equation}
when the same bulk winding is used in both chiral sectors.

For each bipartite EE, the endpoint branch labels are chosen independently of those for the other bipartite entropies.
The independently minimizing choices for three pairs need not extend to a tuple of absolute windings triple.
The branch-dependent factor in the bipartite entropy is therefore
\begin{equation}
    \left[\left| \frac{u_{ij}^2}{u_i' u_j'} \right|^2\right]_{m_{ij}^{(2)}} = \left( \frac{2}{\alpha^2} \right)^2 \left[\calI_{ij}^{\cd}(m_{ij}^{(2)})\right]^2.
\end{equation}

For the tripartite correlator, chose a representative $\bmell$ of $T = [\bmell]$ and set $n_i - n_j = \bar{n}_i - \bar{n}_j = m_{ij}^{(3)}(T)$.
This diagonal choice of holomorphic and antiholomorphic branches is the reality condition for comparison with real bulk geodesic networks.
This relation implies
\begin{equation}
    2 \sin{\frac{\alpha \varphi_{ij}}{2}} \sin{\frac{\alpha \bar{\varphi}_{ij}}{2}} = \cos{(\tau_i - \tau_j)} - \cos{(\vartheta_i - \vartheta_j)} = \calI_{ij}^{\cd}(T).
\end{equation}
where $\varphi_{ij} = \varphi_i - \varphi_j$ and $\bar{\varphi}_{ij} = \bar{\varphi}_i - \bar{\varphi}_j$.
A chosen network branch gives the branch-dependent product
\begin{equation}
    64 \prod_{(ij|X)} \sin{\frac{\alpha \varphi_{ij}}{2}} \sin{\frac{\alpha \bar{\varphi}_{ij}}{2}} = 8 \prod_{(i,j)} \calI_{ij}^{\cd}(T).
    \label{eq_sm:tri-branch-product}
\end{equation}
All factors in \eqref{eq_sm:tri-branch-product} are evaluated on the same absolute winding tuple class.
For the trivalent correlator, this branch is admissible as a real saddle only when the same positivity condition as in the bulk holds,
\begin{equation}
    \calI_{12}^{\cd}(T)>0, \qquad \calI_{23}^{\cd}(T)>0, \qquad \calI_{31}^{\cd}(T)>0,
\end{equation}
and the corresponding square-root branches in the stationary solution are chosen so that $-2P_i\cdot Y>0$ for $i=1,2,3$.
The branch-dependent factor in the tripartite entropy is
\begin{equation}
    \left[\left| \frac{u_{12} u_{23} u_{31}}{u_1' u_2' u_3'} \right|^2\right]_T = \left( \frac{2}{\alpha^2} \right)^3 \calI_{12}^{\cd}(T) \calI_{23}^{\cd}(T) \calI_{31}^{\cd}(T).
\end{equation}

Minimizing each entropy separately, we obtain the vacuum-subtracted tripartite GM as
\begin{equation}
    \varDelta{\gm}^{(3)} = \frac{c}{12} \log{\frac{\displaystyle\min_{T \in \Tadm} \prod_{(i,j)} \calI_{ij}^{\cd}(T)}{\displaystyle\prod_{(ij|X)} \min_{m_{ij}^{(2)} \in \calD_{ij}^{(2)}} \calI_{ij}^{\cd}(m_{ij}^{(2)})}}.
    \label{eq_sm:cft-excess-gme}
\end{equation}
With the Brown--Henneaux relation $c = 3/(2G_N)$~\cite{Brown:1986nw}, this formula matches the backreacted holographic calculation of the vacuum-subtracted tripartite GM in Eq.~\eqref{eq_sm:bulk-excess-gme}.
A nonzero value arises precisely when no admissible common branch induces a tuple that simultaneously minimizes all three bipartite correlators.

\subsection{BTZ continuation}
In the BTZ regime, the same branch-invariant formula follows by analytic continuation $\alpha = \ii \nu$ followed by a separate minimization over positive BTZ branches.
Using $\sin(\ii x)=\ii\sinh x$, the trigonometric branch identities become the hyperbolic formulas after redefining the positive invariant as $\calI_{ij}^{\btz}$.
The branch-independent powers of $\alpha$ become powers of $\nu$ and cancel in the same genuine combination.
The admissible BTZ branches are the family specified above, with $\calI_{ij}^{\btz}>0$ for the RT geodesics and for all three projections of the trivalent network.


\end{document}